\documentclass[preprint,3p,12pt,times,onecolumn]{elsarticle}
\usepackage{amssymb}
\usepackage{amsmath}
\usepackage{graphicx}
\usepackage{bm}
\usepackage{color}
\usepackage[usenames,dvipsnames]{xcolor} % for color changing in edits
\usepackage{ulem} % for \sout
\usepackage{multirow}
\usepackage{natbib}
\usepackage{breqn}
\usepackage{flexisym}
\usepackage{xr}
\usepackage{hyperref}
\usepackage{array} % in preamble
\usepackage{algorithm}
\usepackage{algpseudocode}
\usepackage{enumitem}
\usepackage{upgreek}

\journal{Journal of Computational Physics}

\begin{document}

\begin{frontmatter}



\title{Fast Stokesian Dynamics for Rigid Aggregates} %% Article title


\author{Deepak Mangal$^a$} %% Author name
\author{Avinesh Ojha$^d$}
\author{Wanjiao Liu$^d$}
\author{Ronald G. Larson$^b$}
\author{Jesse Capecelatro$^{a,c}$}
\affiliation{Department of Mechanical Engineering, University of Michigan, Ann Arbor, Michigan 48105, USA}
\affiliation{Department of Chemical Engineering, University of Michigan, Ann Arbor, Michigan 48105, USA}
\affiliation{Department of Aerospace Engineering, University of Michigan, Ann Arbor, Michigan 48105, USA}
\affiliation{Coatings and Surfaces Research, Ford Motor Company, Dearborn, Michigan 48126, USA}

%% Abstract
\begin{abstract}
We present a fast Stokesian dynamics (FSD) framework for the dynamics and rheology of suspensions of rigid aggregates. The method extends the sphere-level formulation of Fiore and Swan [Fiore AM, Swan JW. Fast Stokesian dynamics. J Fluid Mech 2019;878:544–597] to multi-bead rigid bodies. Rigidity is enforced implicitly through geometric constraints, enabling stable and efficient time integration. We develop a block-triangular factorization preconditioner for the resulting saddle-point system. The approach combines an approximate inverse of the far-field mobility with a block-diagonal approximation of the Schur complement, enabling independent inversion of each aggregate sub-block via LU decomposition. The method is implemented as an open-source plugin for the HOOMD-blue software suite. Performance is demonstrated on a range of benchmark problems, including doublet dynamics in shear flow, pair sedimentation, Brownian diffusion, and rheological predictions of suspensions across dilute and structured regimes, with accurate recovery of both deterministic and stochastic behavior. The framework is further validated against experimentally measured rheology of carbon black slurries, explicitly accounting for van der Waals cohesion, Hertzian normal contact, and tangential friction through enhanced lubrication. The simulations accurately reproduce the shear-thinning and high-shear viscous regimes.% but do not capture the low-shear yielding plateau, largely because phase separation prevents access to the homogeneous yielding regime.
The method exhibits favorable GPU scaling for small system sizes, with normalized runtime per bead decreasing before reaching saturation. A size-dependent Ewald splitting parameter accelerates simulations at low volume fractions, yielding up to an order-of-magnitude speedup compared to constant Ewald splitting. For larger systems, a constant Ewald splitting produces linear scaling with particle number, whereas the size-dependent choice leads to quadratic scaling due to increased far-field cost. Overall, the proposed framework enables accurate and scalable simulation of rigid aggregate suspensions in Stokes flow, capturing both dynamics and suspension rheology.%, with further acceleration possible through improved preconditioning and parallelization.
\end{abstract}

\begin{graphicalabstract}
\centering
\includegraphics[width=0.6\textwidth]{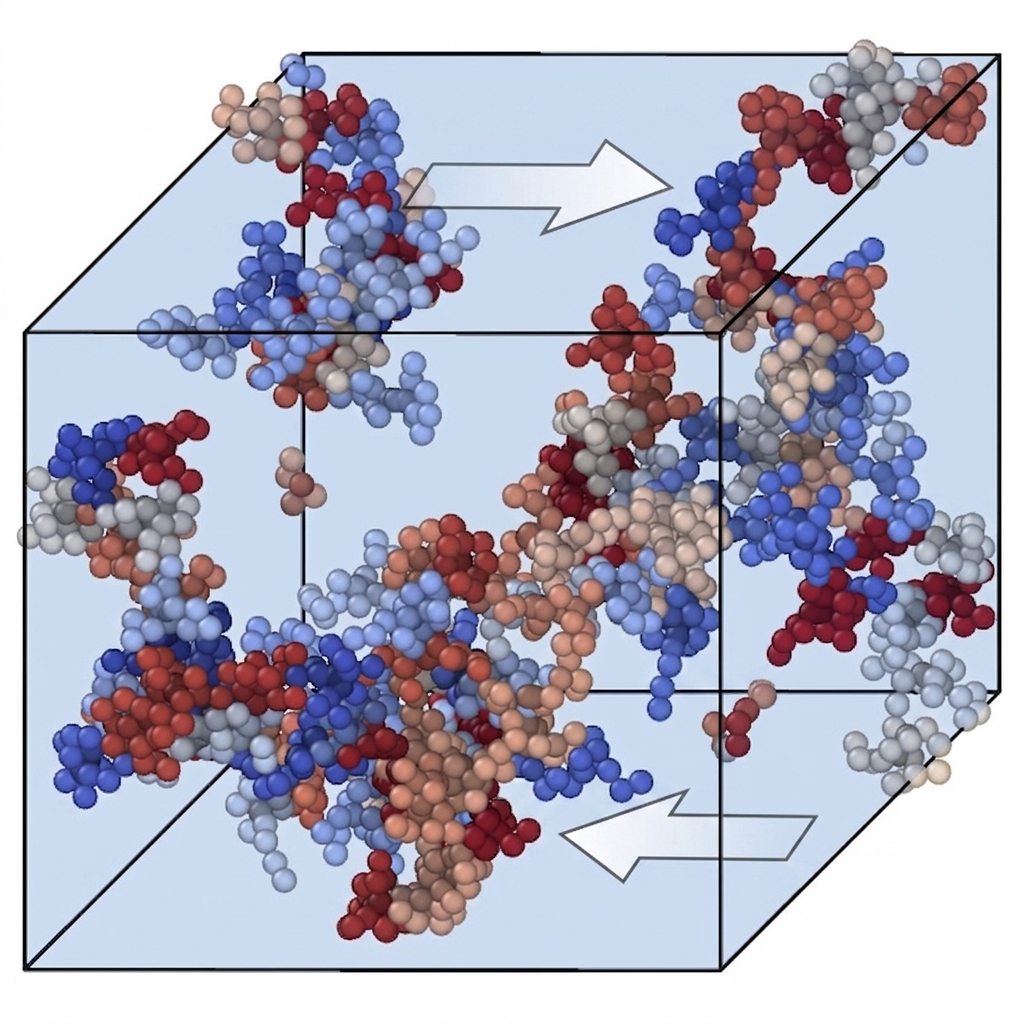}
\end{graphicalabstract}

\begin{highlights}
\item Developed a fast Stokesian dynamics framework for suspensions of rigid aggregates
\item Rigidity enforced implicitly via geometric constraints within a saddle-point system
\item Linear scaling achieved via block-triangular preconditioner
\item Validated against benchmark problems in deterministic and stochastic dynamics
\item Implemented for GPUs within the open-source HOOMD-blue simulation package
\end{highlights}

%% Keywords

\end{frontmatter}



\section{Introduction}
Colloidal suspensions composed of irregular-shaped particles--such as rods, plates, ellipsoids, Janus particles, and fractals aggregates--are encountered in a variety of natural and industrial processes, including liquid crystals, drug delivery, coating, paint, cement, and battery electrode slurries. The macroscopic behavior of such suspensions under flow is governed by microstructural evolution, which manifests through changes in both pair (center-to-center) correlations and particle orientation~\cite{gompper2006soft}. For suspensions of spherical particles, flow-induced changes in macroscopic properties originate solely from modifications to pair correlations. In contrast, non-spherical particles experience hydrodynamic torques that drive alignment and reorientation, leading to anisotropic microstructures that strongly influence bulk rheological response. Rigid aggregates constitute an important subclass of anisotropic systems, characterized by complex geometries formed through the fusion of multiple primary particles. Such permanent rigid aggregates can then reversibly agglomerate into large structures that can form gels, and are commonly encountered in flocculated suspensions, soot dispersions, and battery slurries, where their structure and dynamics critically influence material properties.

At the colloidal scale, particle dynamics in a suspending fluid are governed by hydrodynamic interactions and fluctuating Brownian forces arising from thermal fluctuations, which are coupled through fluctuation--dissipation theorem. Fluid motion in this regime is dominated by viscosity and governed by the Stokes equation, resulting in long-ranged, many-body hydrodynamic coupling between particles. The particle trajectories are typically described by Langevin equations that incorporate both deterministic and stochastic contributions. For rigid aggregates, these dynamics are further constrained by rigid-body motion of the constituent particles, introducing additional coupling between translational and rotational degrees of freedom. Capturing these effects in numerical simulations requires methods that resolve long-ranged, many-body hydrodynamic forces while rigorously enforcing rigid-body constraints within aggregates.

A variety of numerical approaches have been developed to simulate such systems, with particular emphasis on accurately resolving hydrodynamic interactions. These methods broadly fall into two categories: implicit-solvent approaches, such as Stokesian Dynamics (SD), Fast Lubrication Dynamics, and Brownian Dynamics, and explicit-solvent approaches, including Lattice–Boltzmann, Multi-Particle Collision Dynamics, and Dissipative Particle Dynamics~\cite{bolintineanu2014particle}. Among these, Stokesian Dynamics has emerged as one of the most accurate and widely used frameworks for colloidal scale hydrodynamics. 

The classical SD framework, developed by Brady and co-workers~\cite{brady1988stokesian}, captures long-ranged many-body hydrodynamic interactions by decomposing the resistance into a far-field mobility contribution and a pairwise additive near-field correction. This framework has been successfully applied to a broad range of problems, including rheology, sedimentation, diffusion and transport in porous media. However, the method requires inversion of a dense far-field mobility matrix of size $11N \times 11N$ ($N$ being the number of particles), incurring a computational cost of $\mathcal{O}(N^3)$. Consequently, extending these simulations to system sizes relevant to many practical applications remains a significant challenge.

To overcome this limitation, several accelerated formulations of SD have been developed. Sierou and Brady~\cite{sierou2001accelerated} introduced Accelerated Stokesian Dynamics (ASD), which employs a particle–mesh Ewald (PME) approach to evaluate far-field interactions with $\mathcal{O}(N \log{N})$ complexity. Wang and Brady~\cite{wang2016spectral} further improved the efficiency and accuracy of ASD by replacing the PME-based wave-space evaluation with a Spectral Ewald (SE) method. Building on these advances, Fiore et al.~\cite{fiore2017rapid} developed a linearly scaling algorithm for Brownian displacements by decomposing them into independent far-field and near-field contributions, computing the former using fluctuating hydrodynamics with non-uniform FFTs and the latter via Krylov subspace methods. Fiore and Swan~\cite{fiore2019fast} later reformulated SD with stresslet constraints into a saddle-point system, giving rise to the fast Stokesian dynamics (FSD) framework. This framework combines matrix-free operators, GMRES, and tailored preconditioners to achieve $\mathcal{O}(N)$ scaling and has been implemented on GPUs within the HOOMD-blue simulation package~\cite{glaser2015strong} for large-scale simulations of spherical suspensions.

Extending these methods to suspensions of complex-shaped particles poses additional challenges related to particle geometry, kinematic constraints, and multiphysics coupling \cite{delmotte2025modeling}. A common strategy is to represent complex particles as rigid assemblies of spherical beads subject to constraint forces. For instance, Meng and Higdon~\cite{meng2008NonBrownian,meng2008Brownian} studied plate-like particles modeled as rigid clusters of spheres, using the method of reflections for far-field hydrodynamics. More recently, Funkenbusch et al.~\cite{funkenbusch2024approaches} demonstrated that projected conjugate gradient (PrCG) methods can be applied to similar saddle-point systems arising in constrained Brownian dynamics, achieving performance comparable to GMRES while offering an alternative solver strategy. However, their approach is limited to Rotne–Prager–Yamakawa (RPY) hydrodynamics and does not incorporate stresslet constraints or lubrication corrections. While Fiore and Swan~\cite{fiore2019fast} outlined how the FSD framework could, in principle, be extended to rigid bead assemblies, that extension was not implemented or validated, and the associated algorithmic challenges remain unresolved.

This paper presents a fully GPU-accelerated FSD framework for suspensions of rigid aggregates. Section~\ref{sec:method} begins with a brief review of the classical Stokesian dynamics formulation, followed by an overview of the FSD framework for sphere-based suspensions and then its extension to rigid aggregates. %Rigidity is enforced implicitly through geometric constraints embedded within a saddle-point formulation. The resulting system is solved efficiently using block-triangular factorization preconditioner that combines an approximate inverse of the far-field mobility with a block-diagonal approximation of the Schur complement. It enables independent inversion of individual aggregate sub-blocks via LU decomposition. Brownian displacements are computed by decomposing the dynamics into far-field and near-field contributions following the positively split Ewald (PSE) approach. Brownian drift is evaluated using either a random finite-difference estimator or a modified Fixman scheme, as detailed in \ref{app_sec:browniandrift}.
The framework is implemented as an open-source plugin for HOOMD-Blue~\cite{glaser2015strong}. Section~\ref{sec:results} presents validation against benchmark problems for both deterministic and stochastic dynamics, along with a detailed performance analysis. Finally, Sec.~\ref{sec:CB} demonstrates the application of the framework to the rheology and shear-induced microstructural evolution of carbon black slurries.

\section{Simulation method}\label{sec:method}
\subsection{Classical Stokesian dynamics}\label{subsec:classical_sd}
In this section, we briefly review the classical Stokesian Dynamics (SD) method originally developed by Brady and co-workers \cite{brady1988stokesian}. For a suspension of $N$ spherical colloids of radius $a$ dispersed in a Newtonian fluid with viscosity $\eta$, the particle dynamics are governed by the following Langevin equation:
\begin{equation}
    \mathbf{m} \cdot \frac{{\rm d}\mathbf{U}}{{\rm d}t} = \mathbf{F}^{\mathrm{H}} + \mathbf{F}^{\mathrm{B}} + \mathbf{F}^{\mathrm{P}},
    \label{eq:langevin}
\end{equation}
where $\mathbf{m}$ is the $6N \times 6N$ mass/moment-of-inertia tensor, $\mathbf{U}$ is the $6N$ vector of translational and angular velocities, and $\mathbf{F}^{\mathrm{H}}$, $\mathbf{F}^{\mathrm{B}}$, and $\mathbf{F}^{\mathrm{P}}$ denote the hydrodynamic, Brownian, and interparticle forces and torques, respectively. 

In the low particle Reynold number limit ($Re\ll1$), inertial effects are negligible and the fluid motion is governed by the Stokes equations. Under these conditions, the hydrodynamic forces and torques, $\mathbf{F}^{\mathrm{H}}$, and stresslet, $\mathbf{S}^{\mathrm{H}}$, are linearly related to the deviation from the imposed flow field via the grand resistance tensor $\mathbf{R}$, given by
\begin{equation}
    \begin{pmatrix} \mathbf{F}^{\mathrm{H}} \\ \mathbf{S}^{\mathrm{H}}  \end{pmatrix} = 
    \mathbf{R} \cdot \begin{pmatrix} \mathbf{U}^{\infty}-\mathbf{U} \\ \mathbf{E}^{\infty}  \end{pmatrix} =  \begin{pmatrix} \mathbf{R}_{\mathrm{FU}} & \mathbf{R}_{\mathrm{FE}} \\ \mathbf{R}_{\mathrm{SU}} &  \mathbf{R}_{\mathrm{SE}} \end{pmatrix} \cdot \begin{pmatrix} \mathbf{U}^{\infty}-\mathbf{U} \\ \mathbf{E}^{\infty}  \end{pmatrix},
\end{equation}
where $\mathbf{U}^{\infty}$ is the imposed fluid velocity field, and $\mathbf{E}^{\infty}$ is the associated rate of strain tensor evaluated at particle centers. Both hydrodynamic stresslet and strain rate tensors are symmetric, and the trace of the strain rate tensor is zero owing to incompressibility. These tensors can therefore be  represented in a reduced $5N$ vector space, leading to the grand resistance tensor $\mathbf{R}$ of size $11N \times 11N$. The tensor $\mathbf{R}$ is symmetric positive definite and depends only on the particle sizes and their instantaneous positions; its inverse defines the grand mobility tensor.

Thermal fluctuations give rise to the Brownian force $\mathbf{F}^{\mathrm{B}}$, which satisfies the fluctuation--dissipation relation: 
\begin{equation}
    \langle \mathbf{F}^{\mathrm{B}} \rangle = \mathbf{0} 
    \quad \text{and} \quad 
    \langle \mathbf{F}^{\mathrm{B}}(0)\,\mathbf{F}^{\mathrm{B}}(t) \rangle 
    = 2 k_B T \,\mathbf{R}_{\mathrm{FU}}\, \delta(t),
\end{equation}
where angle brackets denote an ensemble average, $k_B$ is the Boltzmann constant, $T$ is the absolute temperature, and $\delta(t)$ is the Direc delta function. The deterministic, non-hydrodynamic force $\mathbf{F}^{\mathrm{P}}$ is arbitrary and may account for any form of interparticle or external force.

Neglecting inertia, the particle velocities $\mathbf{U}$ follow from force balance:
\begin{equation}
     \mathbf{0} = - \mathbf{R}_{\mathrm{FU}} \cdot (\mathbf{U} - \mathbf{U}^{\infty}) + \mathbf{R}_{\mathrm{FE}}:\mathbf{E}^{\infty} + \mathbf{F}^{\mathrm{P}} + \mathbf{F}^{\mathrm{B}},
    \label{eq:euler1}
\end{equation}
which yields
\begin{equation}
    \mathbf{U} = \mathbf{U}^{\infty} + (\mathbf{R}_{\rm FU})^{-1} \cdot (\mathbf{F}^{\mathrm{P}}+\mathbf{R}_{\mathrm{FE}}:\mathbf{E}^{\infty}) + \sqrt{2k_BT/\mathrm{d}{t}} (\mathbf{R}_{\mathrm{FU}})^{-1/2} \cdot \boldsymbol{\psi} + k_B T \nabla \cdot (\mathbf{R}_{\mathrm{FU}})^{-1},
    \label{eq:euler2}
\end{equation}
where the Brownian force $\mathbf{F}^{\mathrm{B}}$ is given by $\sqrt{2k_BT/\mathrm{d}{t}} (\mathbf{R}_{\mathrm{FU}})^{1/2} \cdot \boldsymbol{\psi}$ in accordance with fluctuation--dissipation theorem. Here, $\boldsymbol{\psi}$ is a vector of independent standard Gaussian random variables and $\mathrm{d}{t}$ is the integration timestep size. The last term, $k_B T \nabla \cdot (\mathbf{R}_{\mathrm{FU}})^{-1}$, is the Brownian drift, which is required to generate particle configurations with correct statistics at equilibrium. 

Classical Stokesian dynamics takes advantage of the fact that hydrodynamic interactions among the particles can be decomposed into long-range mobility interactions and short-range lubrication interactions. The long-range interactions are captured through a truncated multipole expansion of the force density together with Faxen's law, yielding the grand mobility tensor $\mathbf{M}^{\infty}$. Its inversion provides an approximation to the far-field resistance. Near-field lubrication interactions are incorporated separately via pairwise resistance contributions $\mathbf{R}_{\mathrm{2B}}$. To avoid double counting, the corresponding far-field pairwise terms $\mathbf{R}_{\mathrm{2B}}^{\infty}$ are subtracted, leading to:

\begin{equation}
    \mathbf{R} = (\mathbf{M}^{\infty})^{-1} + \mathbf{R}_{\mathrm{2B}} - \mathbf{R}_{\mathrm{2B}}^{\infty}.
\end{equation}

Once the grand resistance matrix $\mathbf{R}$ is constructed, particle velocities and configurations can be advanced in time via Eq.~\eqref{eq:euler2}. However, classical SD is computationally demanding: evaluation of $\mathbf{M}^{\infty}$ scales as $\mathcal{O}(N^2)$, while its inversion requires $\mathcal{O}(N^3)$ operations. This limits its applicability to relatively small systems.

\subsection{Fast Stokesian dynamics for suspension of spheres}\label{subsec:fsd_sphere}
Fiore and Swan~\cite{fiore2019fast} developed the fast Stokesian dynamics (FSD) framework to enable large-scale dynamic simulations of colloidal suspensions with full hydrodynamic interactions and Brownian forces. Unlike classical SD, which relies on explicit construction and inversion of dense resistance matrices, FSD reformulates the coupled hydrodynamic problem into a single saddle-point system. This formulation provides a unified and flexible framework for incorporating diverse physical constraints (e.g., rigidity and stresslets), boundary conditions, and complex geometries. The resulting linear systems are solved efficiently using the GMRES iterative solver with a novel preconditioner, eliminating $\mathcal{O}(N^3)$ cost associated with direct matrix inversion. In addition, far-field interactions are accelerated through the particle-mesh-Ewald (PME) method, reducing the overall computational cost to $\mathcal{O}(N\log{N})$. A brief overview of the formulation is presented below; full details can be found in Ref.~\cite{fiore2019fast}.

\subsubsection{Saddle-point formulation}
In classical SD, the resistance matrix, $\mathbf{R}_{\mathrm{FU}}$, is constructed by combining the far-field approximation of mobility $\mathbf{M}^{\infty}$ with near-field lubrication corrections $\mathbf{R}_{\mathrm{FU}}^{\rm nf}$,
\begin{equation}
    \mathbf{R}_{\mathrm{FU}} = \mathcal{B}^T \cdot (\mathbf{M}^{\infty})^{-1} \cdot \mathcal{B} + \mathbf{R}_{\mathrm{FU}}^{\rm nf}.
    \label{eq:Rfu}
\end{equation}
Here, the operator $\mathcal{B}$ maps the motions of rigid particles (translation and rotation) onto generalized local motions, and its adjoint $\mathcal{B}^T$ projects the generalized hydrodynamic force moments back onto the physically relevant particle forces and torques, thereby enforcing the rigidity constraint~\cite{fiore2019fast}:

\begin{equation}
    \begin{pmatrix} \mathbf{U} \\ \mathbf{0} \end{pmatrix} = \mathcal{B} \cdot \mathbf{U}, \quad \mathbf{F} = \mathcal{B}^T \cdot \begin{pmatrix} \mathbf{F} \\ \mathbf{S} \end{pmatrix}.
\end{equation}

For rigid particles in straining flows, the local rate-of-strain must vanish. This condition is expressed by the stresslet constraint~\cite{fiore2018rapid},
\begin{equation}
    \mathbf{M}^{\infty} \cdot \mathcal{F} + \mathcal{B} \cdot \mathbf{U} = 0,
    \label{eq:mob_rel}
\end{equation}
where $\mathcal{F}$ denotes the set of hydrodynamic force moments including the force, torque, and stresslet. Combining this with the force balance Eq.~\eqref{eq:euler1} leads to a coupled saddle-point system for the unknown $\mathcal{F}$ and particle velocities,
\begin{equation}
    \begin{pmatrix} \mathbf{M}^{\infty} & \mathcal{B} \\ \mathcal{B}^T &  -\mathbf{R}_{\mathrm{FU}}^{\rm nf} \end{pmatrix} \cdot \begin{pmatrix} \mathcal{F} \\ \mathbf{U}-\mathbf{U}^{\infty}  \end{pmatrix} = \begin{pmatrix} \begin{pmatrix} \mathbf{0} \\ \mathbf{E}^{\infty}  \end{pmatrix}  \\ -(\mathbf{F}^{\mathrm{P}}+\mathbf{F}^{\mathrm{B}}+\mathbf{R}_{\mathrm{FE}}^{\rm nf}:\mathbf{E}^{\infty})  \end{pmatrix}.
    \label{eq:SP1}
\end{equation}
This formulation avoids explicit construction of the dense resistance matrix and is naturally suited to iterative solvers. 

\subsubsection{Saddle-point preconditioner}
The saddle-point system is efficiently solved using a GMRES solver, and convergence is accelerated using a block preconditioner based on an approximate factorization of the system matrix,
\begin{equation}
    \mathcal{P} = \begin{pmatrix} \mathbf{I} & \mathbf{0} \\ \mathcal{B}^T \cdot (\mathbf{M}^{\infty})^{-1} &  \mathbf{I} \end{pmatrix} \cdot \begin{pmatrix} \mathbf{M}^{\infty} & \mathbf{0} \\ \mathbf{0} &  \mathcal{S} \end{pmatrix} \cdot \begin{pmatrix} \mathbf{I} & (\mathbf{M}^{\infty})^{-1} \cdot \mathcal{B} \\ \mathbf{0} &  \mathbf{I} \end{pmatrix},
    \label{eq:precond}
\end{equation}
where $\mathcal{S}=-(\mathbf{R}_{\mathrm{FU}}^{\rm nf}+\mathcal{B}^T \cdot (\mathbf{M}^{\infty})^{-1} \cdot \mathcal{B})$ is the Schur complement and $\mathbf{I}$ is the identity matrix. To enable efficient application of the preconditioner, the far-field mobility matrix is simplified as $\mathbf{M}^{\infty}=\zeta \mathbf{I}$, where $\zeta$ is an effective drag coefficient. The Schur complement $\mathcal{S}$ is approximated by $\mathcal{S}=\zeta^{-1}\mathbf{I}+\mathbf{\tilde{R}}_{\mathrm{FU}}^{\rm nf}$, where $\mathbf{\tilde{R}}_{\mathrm{FU}}^{\rm nf}$ is a sparse approximation of $\mathbf{R}_{\mathrm{FU}}^{\rm nf}$ built from particle pairs with small separation $r_p<2.1a$. This sparsification significantly reduces both memory cost and computational effort while preserving the dominant lubrication interactions. 

\subsubsection{Stochastic sampling for Brownian motion}
FSD decomposes Brownian motion into far-field and near-field contributions. The far-field contribution is represented as a random slip velocity $\mathbf{U}^{\mathrm{B}}$, computed using the positively split Ewald (PSE) approach. The near-field fluctuations are captured through a Brownian force $\mathbf{F}_{\rm nf}^{\mathrm{B}}$, computed using a fast preconditioned Krylov subspace approximation. The governing saddle-point system then becomes
\begin{equation}
    \begin{pmatrix} \mathbf{M}^{\infty} & \mathcal{B} \\ \mathcal{B}^T &  -\mathbf{R}_{\mathrm{FU}}^{\rm nf} \end{pmatrix} \cdot \begin{pmatrix} \mathcal{F} \\ \mathbf{U}-\mathbf{U}^{\infty}  \end{pmatrix} = \begin{pmatrix} \begin{pmatrix} \mathbf{U}^{\mathrm{B}} \\ \mathbf{E}^{\infty}  \end{pmatrix}  \\ -(\mathbf{F}^P+\mathbf{F}^{\mathrm{B}}_{\rm nf}+\mathbf{R}_{\mathrm{FE}}^{\rm nf}:\mathbf{E}^{\infty})  \end{pmatrix}.
    \label{eq:SP2}
\end{equation}

Within the PSE framework, the random slip velocity $\mathbf{U}^{\mathrm{B}}$ is decomposed into real-space and wave-space components,
\begin{equation}
    \mathbf{U}^{\mathrm{B}} = \mathbf{U}_r^{\mathrm{B}} + \mathbf{U}_w^{\mathrm{B}}.
    \label{eq:UBff}
\end{equation}
The real-space component, $\mathbf{U}_r^{\mathrm{B}}$, is sampled from real-space mobility $\mathbf{M}_r^{\infty}$ as $\mathbf{U}_r^{\mathrm{B}}=\sqrt{2k_B{T}/{\rm d}t}(\mathbf{M}_r^{\infty})^{1/2} \cdot \boldsymbol{\psi}_r$, where $\boldsymbol{\psi}_r$ is a vector of Gaussian random variables. The matrix square root is applied efficiently using an iterative Lanczos scheme~\cite{chow2014preconditioned}. The wave-space contribution, $\mathbf{U}_w^{\mathrm{B}}$, is sampled from wave-space mobility $\mathbf{M}_w^{\infty}$ as $\mathbf{U}_w^{\mathrm{B}}=\sqrt{2k_B{T}/{\rm d}t}(\mathbf{M}_w^{\infty})^{1/2} \cdot \boldsymbol{\psi}_w$. This decomposition enables efficient and scalable sampling of Brownian displacements while remaining consistent with the hydrodynamic mobility. 

Together, the saddle-point formulation, efficient preconditioning, and stochastic sampling strategy enable FSD to achieve near $\mathcal{O}(N)$ scaling for large systems while retaining full hydrodynamic fidelity~\cite{fiore2019fast}.

\subsection{Extension to rigid aggregates}\label{subsec:fsd_agg}
In this work, we extend the FSD framework to rigid aggregates by incorporating geometric constraints that enforce rigid-body motion. This framework enables the efficient simulation of particles with arbitrary shapes, including fractal aggregates, while retaining the matrix-free and scalable structure of FSD. 

Arbitrary shapes are represented by discretizing the particle surface into a set of vertices, with spherical beads placed at these locations. Rigidity can be enforced either explicitly, through stiff bond and angular potentials, or implicitly, through geometric constraints.  Explicit approaches require excessively small timesteps due to the fast relaxation times associated with stiff potentials, which can significantly limit computational efficiency. In contrast, implicit constraint formulations eliminate these fast timescales by directly enforcing rigid-body kinematics, allowing for larger timesteps at the expense of a moderately increased per-step computational cost.

In this work, we adopt the implicit approach to model the dynamics of rigid aggregates. Rather than evolving bead positions independently, the motion of each bead is expressed in terms of the translational and rotational motion of its parent aggregate, thereby enforcing rigidity exactly at the kinematic level.

\subsubsection{Geometric constraint formulation}\label{subsubsec:constraint}
Consider a system of $N$ rigid aggregates comprising a total of $N_{\rm tot}$ beads, with each aggregate containing $n_b=N_{\rm tot}/N$ beads. Let $\mathbf{x} \in \mathbb{R}^{3N_{\rm tot}}$ denote the bead positions and $\mathbf{X} \in \mathbb{R}^{3N}$ the corresponding aggregate centers of mass. The background flow evaluated at bead and aggregate positions is denoted by $\mathbf{u}^{\infty} \in \mathbb{R}^{6N_{\rm tot}}$ and $\mathbf{U}^{\infty}\in\mathbb{R}^{6N}$, respectively. The bead velocity $\mathbf{u}$ and the aggregate velocity $\mathbf{U}$ are related through the geometric constraint~\cite{meng2008NonBrownian}

\begin{equation}
    \mathbf{u} - \mathbf{u} ^{\infty}  = \Sigma^T \cdot  (\mathbf{U} - \mathbf{U}^{\infty}) - \begin{pmatrix} \mathbf{E}^{\infty} \cdot (\mathbf{x} - \mathbf{X}) \\ \mathbf{0} \end{pmatrix},
    \label{eq:gc}
\end{equation}
where $\Sigma^T \in \mathbb{R}^{6N_{\rm tot} \times 6N}$ is the transpose of the geometric constraint matrix that maps aggregate rigid-body motion to bead-level velocities. For bead $i$ belonging to aggregate $j$, the corresponding $6 \times 6$ block of $\Sigma$ is

\begin{equation}
    \Sigma_{i,j} = \begin{pmatrix} \mathbf{I}_3 & \mathbf{0} \\ \boldsymbol{\epsilon} \cdot (\mathbf{x}_i - \mathbf{X}_j) & \mathbf{I}_3 \end{pmatrix},
    \label{eq:sigma}
\end{equation}
where $\boldsymbol{\epsilon}$ is the Levi-Civita tensor and $\mathbf{I}_3 \in \mathbb{R}^{3 \times 3}$ is the identity matrix. This construction ensures that all beads within an aggregate undergo a consistent rigid-body motion composed of translation and rotation.

\subsubsection{Hydrodynamic and inter-particle forces}\label{subsubsec:forces}
The hydrodynamic interactions $\mathbb{F}^{\mathrm{H}}$ acting on the aggregates are decomposed into far-field and near-field contribution,
\begin{equation}
    \mathbb{F}^{\mathrm{H}} = \mathbb{F}_{\rm ff}^{\mathrm{H}} + \mathbb{F}_{\rm nf}^{\mathrm{H}} =  \Sigma \cdot \mathcal{B}^T \cdot \mathcal{F} - \mathbb{R}_{\mathrm{FU}}^{\rm nf} \cdot (\mathbf{U}- \mathbf{U}^{\infty})  + \mathbb{R}_{\mathrm{FE}}^{\rm nf}:\mathbb{E}^{\infty},
    \label{eq:FH_agg}
\end{equation}
where the near-field resistance tensor for the aggregate is given by $\mathbb{R}^{\rm nf}=\Sigma \cdot \mathbf{R}^{\rm nf} \cdot \Sigma^T$. The Brownian forces $\mathbb{F}^B$ are similarly decomposed as

\begin{equation}
    \mathbb{F}^{\mathrm{B}} = \mathbb{F}_{\rm ff}^{\mathrm{B}} + \mathbb{F}_{\rm nf}^{\mathrm{B}} =  \Sigma \cdot \mathcal{B}^T \cdot (\mathbf{M}^{\infty})^{-1} \cdot \mathbf{u}^{\mathrm{B}} + \Sigma \cdot \mathbf{F}_{\rm nf}^{\mathrm{B}},
    \label{eq:FB_agg}
\end{equation}
where $\mathbf{M}^{\infty}$ is the far-field mobility matrix, $\mathbf{u}^{\mathrm{B}}$ is the far-field Brownian velocity, and $\mathbf{F}_{\rm nf}^{\mathrm{B}}$ is near-field Brownian force. The conservative forces are accumulated from bead-level interactions as

\begin{equation}
\mathbb{F}^{\mathrm{P}} = \Sigma \cdot \mathbf{F}^{\mathrm{P}}.
\end{equation}

\subsubsection{Saddle-point formulation and preconditioner}\label{subsubsec:SPF}
Combining the bead-level mobility relation~\eqref{eq:mob_rel} with the aggregate-level force balance~\eqref{eq:euler1} yields the following saddle-point problem for aggregate velocities:
\begin{equation}
    \begin{pmatrix} \mathbf{M}^{\infty} & \mathcal{B} \cdot \Sigma^T \\ \Sigma \cdot \mathcal{B}^T &  -\mathbb{R}_{\mathrm{FU}}^{\rm nf} \end{pmatrix} \cdot \begin{pmatrix} \mathcal{F} \\ \mathbf{U}- \mathbf{U}^{\infty}  \end{pmatrix} = \begin{pmatrix} \begin{pmatrix} \mathbf{E}^{\infty} \cdot (\mathbf{x}-\mathbf{X}) + \mathbf{u}^{\mathrm{B}} \\ \mathbf{E}^{\infty}  \end{pmatrix}  \\ -(\mathbb{F}^P+\mathbb{F}_{\rm nf}^{\mathrm{B}}+\mathbb{R}_{\mathrm{FE}}^{\rm nf}:\mathbb{E}^{\infty})  \end{pmatrix}.
    \label{eq:spm_agg}
\end{equation}

To solve this system efficiently, we employ a block-triangular preconditioner analogous to that used in the sphere-based FSD formulation (shown in Table~\ref{tab:fsd}). The far-field mobility block is approximated as $\mathbf{M}^{\infty}\approx\zeta\mathbf{I}$, where $\zeta=6\pi\eta a$, and the Schur complement is approximated by block-diagonal operator,

\begin{equation}
    \mathcal{S}\approx \text{blockdiag}[-(\Sigma \cdot (\zeta^{-1}\mathbf{I}+\tilde{\mathbf{R}}_{\mathrm{FU}}^{\rm nf}) \cdot \Sigma^T)]^{-1},
\end{equation}
where $\tilde{\mathbf{R}}_{\mathrm{FU}}^{\rm nf}$ is constructed using only bead pairs with separation $r<2.1a$. In the Schur complement, only the diagonal blocks associated with individual rigid aggregates are inverted independently and used for preconditioning. This avoids forming or storing the full off-diagonal coupling structure while preserving the dominant block-diagonal terms for each aggregate. The resulting preconditioner significantly improves the conditioning of the system and accelerates Krylov subspace convergence without incurring the cost of exact factorization.

\begin{table}[h!]
\centering
\renewcommand{\arraystretch}{2}
\begin{tabular}{>{\centering\arraybackslash}m{2cm}| >{\centering\arraybackslash}m{6cm}|>{\centering\arraybackslash}m{6cm}}
& \textbf{Spheres} & \textbf{Aggregates} \\
\hline
Saddle point matrix &
$
\begin{aligned}
\begin{pmatrix} 
\mathbf{M}^{\infty} & \mathcal{B} \\ 
\mathcal{B}^T & -\mathbf{R}_{\mathrm{FU}}^{\rm nf} 
\end{pmatrix} 
\cdot 
\begin{pmatrix} 
\mathcal{F} \\ 
\mathbf{U}-\mathbf{U}^{\infty} 
\end{pmatrix} 
&= \\
\begin{pmatrix} 
\begin{pmatrix} \mathbf{0} \\ \mathbf{E}^{\infty} \end{pmatrix} \\
-(\mathbf{F}^{\mathrm{P}} + \mathbf{F}^{\mathrm{B}} + \mathbf{R}_{\mathrm{FE}}^{\rm nf} : \mathbf{E}^{\infty}) 
\end{pmatrix}
\end{aligned}
$
&
$
\begin{aligned}
\begin{pmatrix} 
\mathbf{M}^{\infty} & \mathcal{B} \cdot \Sigma^T \\ 
\Sigma \cdot \mathcal{B}^T & -\mathbb{R}_{\mathrm{FU}}^{\rm nf} 
\end{pmatrix} 
\cdot 
\begin{pmatrix} 
\mathcal{F} \\ 
\mathbf{U}-\mathbf{U}^{\infty} 
\end{pmatrix} 
&=\\ 
\begin{pmatrix} 
\begin{pmatrix} \mathbf{E}^{\infty}\cdot(\mathbf{X}-\mathbf{x}) \\ \mathbf{E}^{\infty} \end{pmatrix} \\
-(\mathbb{F}^{\mathrm{P}} + \mathbb{F}^{\mathrm{B}} + \mathbb{R}_{\mathrm{FE}}^{\rm nf} : \mathbb{E}^{\infty}) 
\end{pmatrix}
\end{aligned}
$
\\
\hline
Precondi- tioner & 
$
\begin{aligned}
\mathcal{P}^{-1} =
\begin{pmatrix} 
\mathbf{I} & -\mathcal{B} \\ 
\mathbf{0} & \mathbf{I} 
\end{pmatrix}
\\
\cdot
\begin{pmatrix} 
\zeta \mathbf{I} & -\mathbf{0} \\ 
\mathbf{0} & -(\mathbf{I}+\tilde{\mathbf{R}}_{\mathrm{FU}}^{\rm nf})^{-1} 
\end{pmatrix}
\cdot
\begin{pmatrix} 
\mathbf{I} & \mathbf{0} \\ 
-\mathcal{B}^T & \mathbf{I} 
\end{pmatrix}
\end{aligned}
$
&
$
\begin{aligned}
\mathcal{P}^{-1} =
\begin{pmatrix} 
\mathbf{I} & -\mathcal{B} \cdot \Sigma^T \\ 
\mathbf{0} & \mathbf{I} 
\end{pmatrix}
\\
\cdot
\begin{pmatrix} 
\zeta \mathbf{I} & -\mathbf{0} \\ 
\mathbf{0} & -(\Sigma \cdot (\mathbf{I}+\tilde{\mathbf{R}}_{\mathrm{FU}}^{\rm nf}) \cdot \Sigma^T) ^{-1} 
\end{pmatrix}
\cdot
\begin{pmatrix} 
\mathbf{I} & \mathbf{0} \\ 
-\Sigma \cdot \mathcal{B}^T & \mathbf{I} 
\end{pmatrix}
\end{aligned}
$
\end{tabular}
\caption{Saddle point formulation and preconditioner for colloidal sphere and aggregate suspensions.}
\label{tab:fsd}
\end{table}

\subsubsection{Brownian drift}\label{subsubsec:drift}
To ensure consistency with the Gibbs–Boltzmann distribution in the presence of configuration-dependent mobility, the Brownian drift term in Eq.~\eqref{eq:euler2} must be included. In the FSD framework, this term is evaluated separately and added to the particle velocities obtained from the saddle-point solve (Eq.~\eqref{eq:SP2}). In this work, we consider two established approaches for evaluating this term: the Random Finite Difference (RFD) method~\cite{delong2014brownian} and a modified Fixman scheme~\cite{fixman1978simulation}.

The RFD method approximates the divergence of the mobility using a centered finite-difference based on small random perturbations of the particle configuration. This requires two additional saddle-point solves at perturbed states to estimate the drift, with the perturbation size chosen to balance truncation and iterative solver errors.

The modified Fixman scheme instead incorporates the drift implicitly through a midpoint discretization. Brownian increments are first generated using the mobility at the current configuration, and two saddle-point solves are then used to recover the drift from differences in the resulting velocities. This approach couples drift evaluation with Brownian sampling, reducing the total computational cost.

In the original FSD implementation, RFD required four saddle-point solves per time step: two for drift evaluation, one for Brownian displacement, and one for the deterministic update. The modified Fixman scheme reduces this to three solves by combining drift evaluation with Brownian sampling. Both approaches were validated using a system of two particles interacting through a linear potential, yielding equilibrium distributions in excellent agreement with theoretical predictions. Additional details and quantitative comparisons are provided in ~\ref{app_sec:browniandrift}.

\subsection{Time integration}\label{subsubsec:integration}
Once the translational and angular velocities of the aggregates are obtained from the saddle-point solve, the aggregate configurations are advanced in time using a stochastic Euler–Maruyama scheme. The center-of-mass position is updated explicitly using the computed translational velocity as
\begin{equation}
    \mathbf{X}(t+\mathrm{d}{t}) = \mathbf{X}(t) + \mathbf{U} \mathrm{d}{t}.\\
    \label{eq:EM1}
\end{equation}

The orientations are represented using quaternions to ensure numerical stability and avoid singularities associated with other parametrizations (e.g., Euler angles). The quaternion $\mathbf{q}_c$ is updated using the angular velocity $\mathbf{\Omega}$ as

\begin{equation}
    \mathbf{q}_c(t+\mathrm{d}{t}) = \mathbf{q}_c(t) + \frac{\mathrm{d}{t}}{2} \mathbf{\Omega} \times \mathbf{q}_c(t),
    \label{eq:EM2}
\end{equation}
followed by normalization to maintain unit magnitude.

Given the updated center-of-mass position and orientation, bead positions are reconstructed by enforcing rigid-body kinematics:

\begin{equation}
    \mathbf{x}(t+\mathrm{d}{t}) = \mathbf{x}(t) + \mathbf{q}_c \mathbf{x}_{\rm body} \mathbf{q}_c^{-1},
    \label{eq:EM3}
\end{equation}
where $\mathbf{x}_{\rm body}$ denotes the bead position in the body-fixed frame. This reconstruction ensures that the internal geometry of each aggregate remains strictly rigid throughout the simulation.

\subsection{Numerical implementation}\label{subsec:Numerics}
The proposed rigid-aggregate FSD framework is implemented as an open-source plugin within the HOOMD-blue software suite~\cite{glaser2015strong}, extending the original FSD implementation for unconstrained spheres~\cite{fiore2019fast}. HOOMD-blue is a versatile particle simulation platform designed for nano- and colloidal-scale molecular dynamics and hard particle Monte Carlo simulations. The framework presented here leverages HOOMD-blue's Python interface and GPU-accelerated C++/CUDA backend to enable large-scale simulations. All operators are implemented in a matrix-free manner, such that only matrix–vector products are evaluated. This avoids explicit assembly and storage of large dense mobility or resistance matrices, which is critical for achieving scalability on modern GPU architectures.

For clarity and reproducibility, the complete sequence of operations in the rigid-aggregate FSD framework is summarized in Algorithm~\ref{alg:fsd}, which outlines the implementation details and computational workflow used in this work.

\clearpage

\begin{algorithm}[H]
\small % or \footnotesize
\caption{FSD algorithm for rigid aggregates}\label{alg:fsd}
\begin{algorithmic}[]
\Statex \textbf{Step 1: Construct preconditioner}
\begin{itemize}[noitemsep, topsep=0pt]
    \item Construct the bead–level sparse lubrication resistance matrix \(\tilde{\mathbf{R}}_{\mathrm{FU}}^{\rm nf}\) using a reduced cut-off $r_p=2.1a$ in CSR (compressed sparse row) format. Only interactions between beads on \textbf{different bodies} are included.
    \item Compute rigid-body level matrix using 
    \[
    \tilde{\mathbb{R}}_{\mathrm{FU}}^{\rm nf} = \Sigma \cdot (\mathbf{I} + \tilde{\mathbf{R}}_{\mathrm{FU}}^{\rm nf}) \cdot \Sigma^T,\
    \]
    where \(\Sigma\) is the bead-to-body mapping matrix. The resulting matrix \(\tilde{\mathbb{R}}_{\mathrm{FU}}^{\rm nf}\) is block diagonal, with each diagonal block corresponding to one rigid body.
    \item For each body $i$, compute and store the inverse of diagonal block $(\Sigma \cdot (\mathbf{I} + \tilde{\mathbf{R}}_{\mathrm{FU}}^{\rm nf}) \cdot \Sigma^T)_{ii}^{-1}$ using LU factorization, which will serve as the local preconditioner for GMRES iterations.
\end{itemize}
\Statex \textbf{Step 2: Evaluate non-hydrodynamic forces} 
\begin{itemize}[noitemsep, topsep=0pt]
    \item Compute bead-level non-hydrodynamic forces $\mathbf{F}^{\mathrm{P}}$, including Derjaguin, Landau, Verwey, and Overbeek (DLVO) interactions and normal elastic forces, considering only inter-body bead interactions (Sec.~\ref{sec:interact}).
    \item Project these forces to the rigid-body level using  \(\mathbb{F}^{\mathrm{P}} = \Sigma \cdot \mathbf{F}^{\mathrm{P}}\)
\end{itemize}
\Statex \textbf{Step 3: Compute Brownian contribution}
\begin{itemize}[noitemsep, topsep=0pt]
    \item \textbf{Far-field Brownian velocity}
    \begin{itemize}[noitemsep, topsep=0pt]
        \item Far-field velocity is decomposed into real-space and wave-space contributions, corresponding to the Ewald sum
        \[
        \mathbf{u}^{\mathrm{B}} = \mathbf{u}_r^{\mathrm{B}} + \mathbf{u}_w^{\mathrm{B}}.
        \]
        \item The real-space contribution \(\mathbf{u}_r^{\mathrm{B}}\) is sampled from \(\mathbf{M}_r^{\infty}\) using iterative Lanczos technique \(\mathbf{u}_r^{\mathrm{B}} = \sqrt{\frac{2k_B{T}}{{\rm d}t}}(\mathbf{M}_r^{\infty})^{1/2} \cdot \boldsymbol{\psi}_r\),        where \(\boldsymbol{\psi}_r\) is a vector of independent standard Gaussian random numbers.
        \item Wave-space contribution is sampled from \(\mathbf{M}_w^{\infty}\), whose square root is analytically available, via \(\mathbf{u}_w^{\mathrm{B}} = \sqrt{\frac{2k_B{T}}{{\rm d}t}}(\mathbf{M}_w^{\infty})^{1/2} \cdot \boldsymbol{\psi}_w\) with \(\boldsymbol{\psi}_w\) being independent Gaussian noise.
    \end{itemize}
    \item \textbf{Near-field Brownian force}
    \begin{itemize}[noitemsep, topsep=0pt]
        \item The bead-level near-field Brownian force \(\mathbf{F}_{\rm nf}^{\mathrm{B}}\) is sampled from \(\mathbf{R}_{\mathrm{FU}}^{\rm nf}\) using Lanczos technique \(\mathbf{F}_{\rm nf}^B = \sqrt{\frac{2k_B{T}}{{\rm d}t}}(\mathbf{R}_{\mathrm{FU}}^{\rm nf})^{1/2} \cdot \boldsymbol{\psi}_{\rm nf}\),        where \(\boldsymbol{\psi}_{\rm nf}\) is a vector of independent standard Gaussian random numbers.
        \item Project these forces to the rigid-body level using  \(\mathbb{F}_{\rm nf}^{\mathrm{B}} = \Sigma \cdot \mathbf{F}_{\rm nf}^{\mathrm{B}}\).
    \end{itemize}
    \item \textbf{Brownian Drift term (RFD or modified Fixman scheme)} - see \ref{app_sec:browniandrift}
\end{itemize}
\Statex \textbf{Step 4: Solve for rigid-body velocities}
\begin{itemize}[noitemsep, topsep=0pt]
    \item Use GMRES with the local preconditioner from Step 1 to solve the saddle-point matrix problem (Eq.~\eqref{eq:spm_agg}) and obtain rigid-body velocities.
\end{itemize}
\Statex \textbf{Step 5: Update rigid-body configuration}
\begin{itemize}[noitemsep, topsep=0pt]
    \item Advance positions and orientations of all rigid bodies using the computed velocities (See Sec.\ref{subsubsec:integration}).
    \item Update bead positions within each rigid body via the geometric constraint (Eq.~\eqref{eq:EM3}).
\end{itemize}
\end{algorithmic}
\end{algorithm}
\clearpage

\section{Results \& discussion}\label{sec:results}
We validate the framework by comparing its predictions against established static and dynamic results across a range of benchmark problems. We first assess the deterministic component by simulating the motion of a single doublet under steady shear and the sedimentation of two interacting spheres. The stochastic component is then examined through short-time translational and rotational diffusivities of suspensions composed of randomly distributed aggregate spheres. Finally, we evaluate stress predictions via shear/extensional viscosity in ordered and dilute random systems, and demonstrate computational efficiency through benchmarking in random fiber suspensions over a range of volume fractions, $\phi$.

\subsection{Validation}\label{subsec:validation}
\subsubsection{Single doublet under steady shear}
In order to validate the deterministic formulation, we first consider a doublet of equal-sized, contacting spheres positioned on the $x-y$ plane and subjected to a simple shear flow, with flow in $x$-direction and velocity gradient along $y$-direction. The angular velocity and hydrodynamic forces are computed as functions of the orientation angle $\theta$ relative to the shear gradient direction. 

\begin{figure}
\centering
\includegraphics[scale=1]{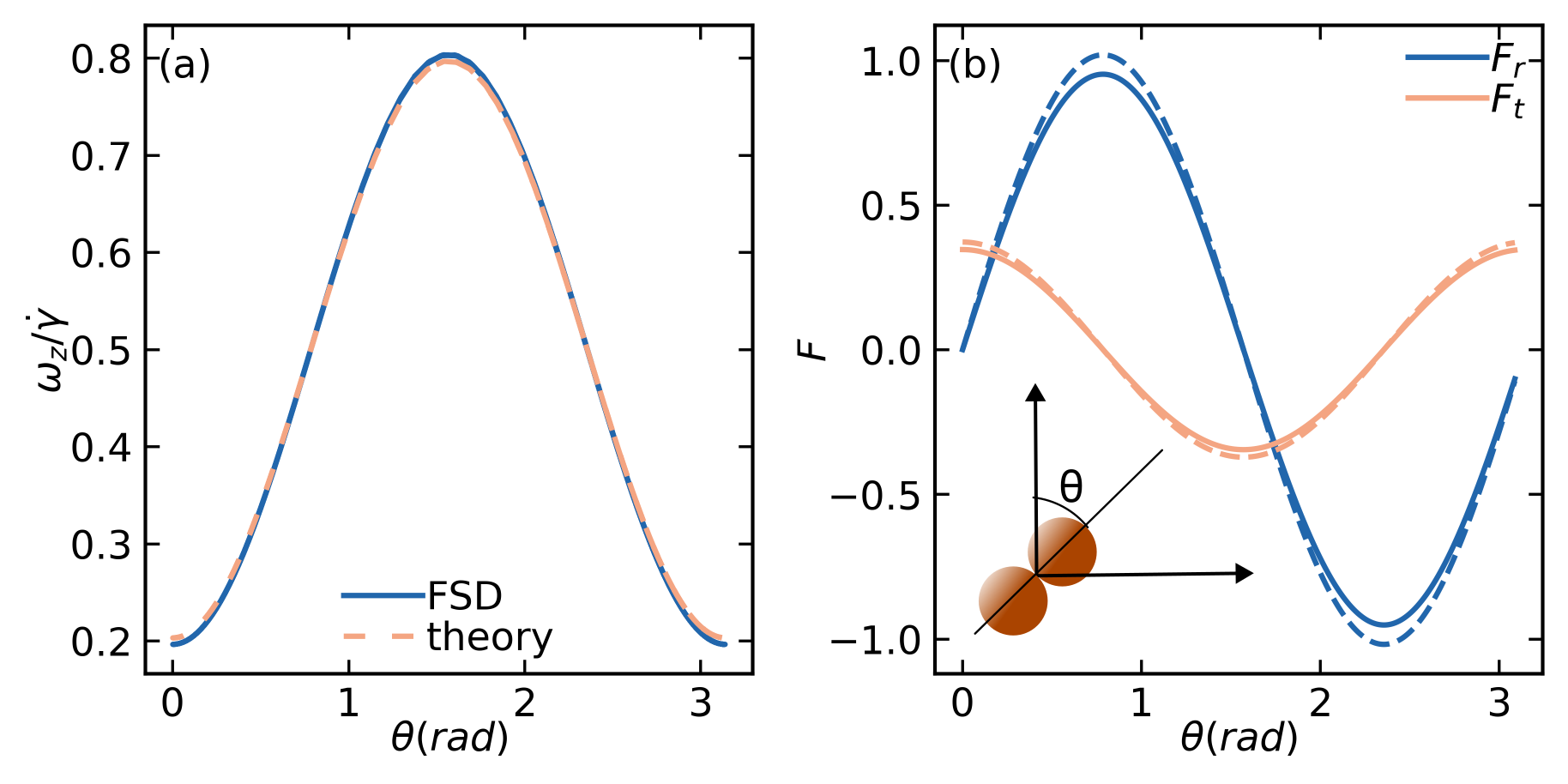}
\caption{(a) Angular velocity and (b) hydrodynamic forces of doublet versus orientation in steady shear.}
\label{fig:doublet}
\end{figure}

As shown in Fig.~\ref{fig:doublet}(a), the predicted normalized angular velocity $\omega_z/\dot{\gamma}$ agrees well with analytical solution $\omega_z/\dot{\gamma} =  - 0.5 (1+0.594 \cos2\theta)$~\cite{nir1973creeping}. The radial and tangential components of the hydrodynamic force acting on either sphere of the doublet are 
\begin{equation}
\begin{aligned}
    F_r(\theta) &= \frac{\pi \eta a^2}{2} \dot{\gamma} (h_1 + h_2) \sin(2\theta) \quad\text{and}\quad
    F_t(\theta) &= \frac{\pi \eta a^2}{2} \dot{\gamma} h_1 \cos(2\theta),
\end{aligned}
\end{equation}
where $h_1$ and $h_2$ are 4.463 and 7.767, respectively. Figure~\ref{fig:doublet}(b) presents the radial and tangential component of hydrodynamic forces. The radial component matches the analytical solution with high accuracy, while the tangential component is slightly underestimated. This discrepancy is likely due to the truncated multipole expansion inherent to the Stokesian Dynamics formulation, which limits the resolution of higher-order and tangential hydrodynamic interactions.

\subsubsection{Sedimentation of two spheres}
We next examine the sedimentation of two identical aggregate spheres under gravity in a periodic box of size $L=150d$. Each aggregate sphere is discretized using $n_b=100$ beads. The settling velocity is computed as a function of the center-to-center separation for both horizontal and vertical configurations, and normalized by the velocity of an isolated sphere in the same domain. The results are compared with those from an equivalent sphere-level model. 

\begin{figure}
\centering
\includegraphics[scale=1]{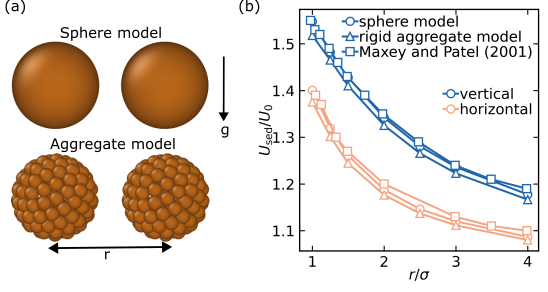}
\caption{(a) Sphere and rigid-aggregate models of two spheres settling under gravity in a periodic domain with $L/\sigma=150$. (b) Scaled settling velocity as a function of center-to-center distance in the vertical and horizontal directions, comparing with sphere-level model and prior results from Maxey \& Patel (2001). $U_0$ is the settling of isolated sphere in the same periodic domain.}
\label{fig:pair_uset}
\end{figure}

Figure~\ref{fig:pair_uset}(a) shows sphere-level and aggregate models of the two spheres. Consistent with the symmetry of Stokes flow, both particles exhibit identical settling velocities. For a fixed separation, the vertical configuration yields higher settling velocities than the horizontal configuration (Fig.~\ref{fig:pair_uset}(b)). Overall, the results show strong agreement with the sphere-level model and prior work by Maxey \& Patel~\cite{maxey2001localized}.

\subsubsection{Concentrated random suspension of rigid hard-spheres}
We next validate the stochastic component of the framework using Brownian dynamics simulation of rigid hard-spheres in a periodic domain of size $L=100$. Each aggregate sphere is discretized using $n_b=42$ beads. To avoid unphysical overlap between beads belonging to different spheres, a normal contact potential with stiffness $k=10^4k_B{T}$ is applied. The short-time translational and rotational diffusion coefficients are computed from stochastic velocity using

\begin{equation}
    D = \delta t \langle \mathbf{U} \cdot \mathbf{U} \rangle,
\end{equation}
where $\mathbf{U}$ denotes the instantaneous stochastic velocity. Both diffusivities are normalized by those of an isolated sphere with the same hydrodynamic radius $R_{\mathrm{h}}$.

The effective hydrodynamic radii $R_{\mathrm{h}}$ are obtained from separate simulations of settling of an isolated sphere under an applied force $f_p=1$ or torque $\tau_p$, given by \cite{vazquez2014multiblob}
\begin{equation}
\begin{aligned}
    R_{\mathrm{h},f} &= \frac{f_p}{6\pi\eta u_p} \qquad
    R_{{\mathrm{h}},\tau} &= \frac{\tau_p}{6\pi\eta \omega_p},
\end{aligned}
\end{equation}
where $u_p$ and $\omega_p$ are the resulting translational and angular velocities of the sphere, respectively. A finite-size correction is then applied to extrapolate the hydrodynamic radius to the infinite-domain limit ($L \to \infty$):
\begin{equation}
    R_{\mathrm{h}}(\infty)=\frac{R_{\mathrm{h}}(L)}{1+2.84R_{\mathrm{h}}(L)/L}.
\end{equation}

\begin{figure}
\centering
\includegraphics[scale=0.75]{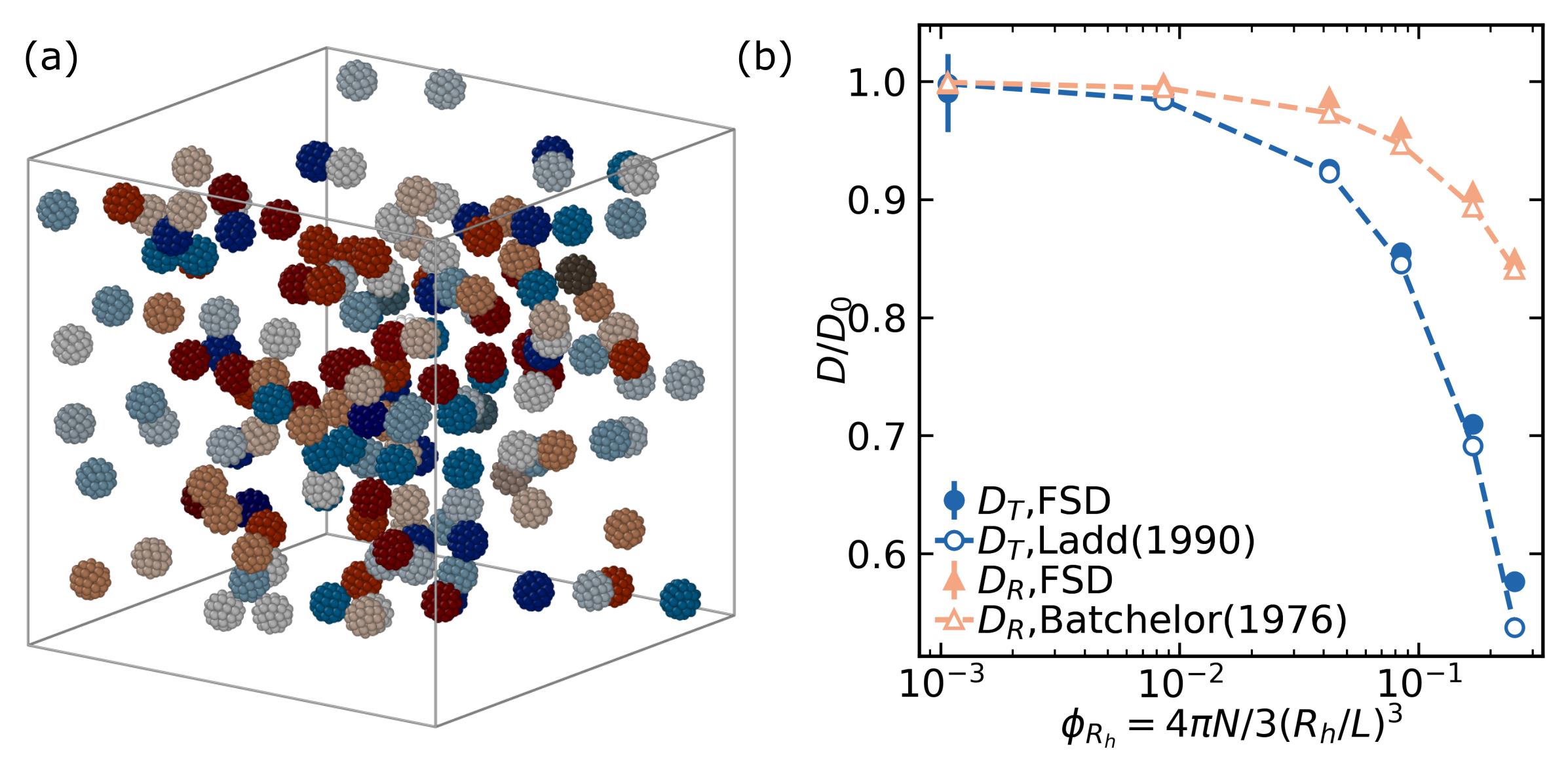}
\caption{(a) Snapshot of a random suspension of rigid hard spheres at volume fraction $\phi=0.1$. Each sphere is represented by $n_b = 42$ beads uniformly distributed on a spherical surface of diameter $d = 6$. (b) Short-time translational and rotational diffusion coefficients as a function of volume fraction $\phi$ for a system of size $L=100$. Both coefficients are computed from stochastic displacements using $D = (\delta t / 6 k_B T)\langle \mathbf{U} \cdot \mathbf{U} \rangle$ and are normalized by the translational and rotational diffusivity of an isolated sphere with the same hydrodynamic radius $R_h$. The translational diffusivity, corrected for finite-size effects, is compared with the results of Ladd~\cite{ladd1990hydrodynamic}, while the rotational diffusivity is compared with Batchelor’s theoretical predictions~\cite{batchelor1976brownian}.}
\label{fig:diffusion}
\end{figure}

Figure~\ref{fig:diffusion}(a) shows a representative snapshot of a random suspension of rigid hard spheres at volume fraction $\phi=0.1$. Figure~\ref{fig:diffusion}(b) shows the short-time diffusivity as function of sphere volume fraction $\phi$, along with analytical predictions. The translational diffusivity, corrected for finite-size effects, agrees well with the results of Ladd~\cite{ladd1990hydrodynamic}, while the rotational diffusivity is consistent with Batchelor’s theoretical predictions~\cite{batchelor1976brownian}. Minor deviations are observed at higher volume fractions, with more pronounced discrepancies in rotational diffusivity. These differences are primarily attributed to the discretized representation of the aggregate sphere, which introduces surface roughness and reduces hydrodynamic resistance compared to ideal smooth spheres, particularly affecting rotational motion.

\subsubsection{Stress calculations for different systems}
We finally validate the stress evaluation by computing viscosity of different aggregate systems. The volume-averaged stress for a system containing $N$ aggregates in volume $V$ is given by

\begin{equation}
    \langle\boldsymbol{\sigma}\rangle = -n k_BT\mathbf{I} + 2\eta \mathbf{E}^{\infty} + \frac{N}{V}(\langle \mathbf{S}^{\mathrm{H}} \rangle + \langle \mathbf{S}^{\mathrm{B}} \rangle + \langle \mathbf{S}^{\mathrm{P}} \rangle + \langle \mathbf{S}^{\mathrm{C}} \rangle),
\end{equation}
where $n$ is the number density. The first term represents the isotropic Brownian contribution, while the second corresponds to the Newtonian solvent stress.

The particles contribution to the stress consists of four components: the hydrodynamic stress $\langle \mathbf{S}^{\mathrm{H}} \rangle$, the Brownian stress $\langle \mathbf{S}^{\mathrm{B}} \rangle$, the inter-particle (elastic) $\langle \mathbf{S}^{\mathrm{P}} \rangle$, and constrained stress $\langle \mathbf{S}^{\mathrm{C}} \rangle$, which arises due to the non-uniform force distribution within rigid aggregates. These contributions are defined as

\begin{subequations}\label{eq:stress_comp}
\begin{align}
    \langle \mathbf{S}^{\mathrm{H}} \rangle &=\langle \mathbf{S}^{\rm ff}-\mathbf{R}_{\mathrm{SU}}^{\rm nf}\cdot(\mathbf{u}-\mathbf{u}^{\infty})+\mathbf{R}_{\mathrm{SE}}^{\rm nf}:\mathbf{E}^{\infty} \rangle\\
    \langle \mathbf{S}^{\mathrm{B}} \rangle &= -k_{B}T\langle \nabla \cdot (\mathbf{R}_{\mathrm{SU}}\cdot \mathbf{R}_{\mathrm{FU}}^{-1}) \rangle\\
    \langle \mathbf{S}^{\mathrm{P}} \rangle &= -\langle\mathbf{x}\mathbf{F}^{\mathrm{P}}\rangle\\
    \langle \mathbf{S}^{\mathrm{C}} \rangle &= -\frac{1}{2}\langle(\mathbf{x}-\mathbf{X})\otimes\mathbf{F}^{\mathrm{C}} + ((\mathbf{x}-\mathbf{X})\otimes\mathbf{F}^{\mathrm{C}})^T\rangle,
\end{align}
\end{subequations}
where $\mathbf{R}_{\mathrm{SU}}$ and $\mathbf{R}_{\mathrm{SE}}$ are the particle configuration dependent resistance tensors relating the stresslet to the particle velocity and imposed rate of strain $\mathbf{E}^{\infty}$, respectively. The constrained force $\mathbf{F}^{\mathrm{C}}$ is the negative sum of all forces acting on each bead.

\begin{figure}
\centering
\includegraphics[scale=1]{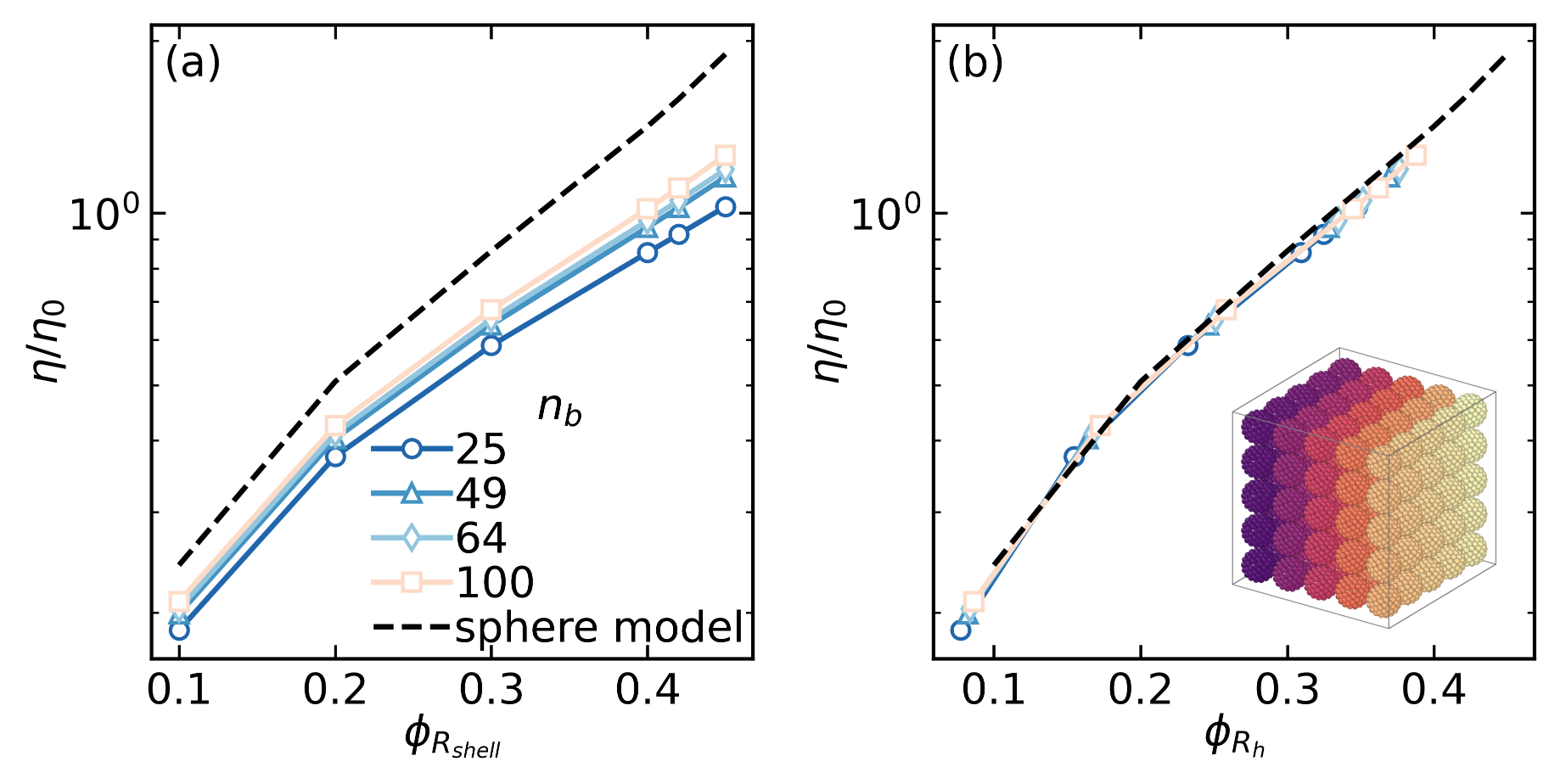}
\label{fig:lattice_stress}
\end{figure}

We first consider simple cubic lattices of aggregate spheres and compute the shear viscosity as a function of volume fraction. Each aggregate sphere (with shell radius $r_{\mathrm{shell}}=1$) is discretized using different numbers of beads $n_b$. Figure~\ref{fig:lattice_stress}(a) shows the viscosity for different $n_b$ along with the sphere-level model. When the volume fraction is defined based on $r_{\mathrm{shell}}$, the aggregate results deviate from the sphere-level predictions. However, when the volume fraction is defined based on the aggregate hydrodynamic radius, the data for all $n_b$ collapse onto a single curve and agree well with the sphere-level model (Fig.~\ref{fig:lattice_stress}(b)). This demonstrates that the hydrodynamic radius provides a consistent measure for comparing discretized aggregates with continuum spheres.

\begin{figure}
\centering
\includegraphics[scale=1]{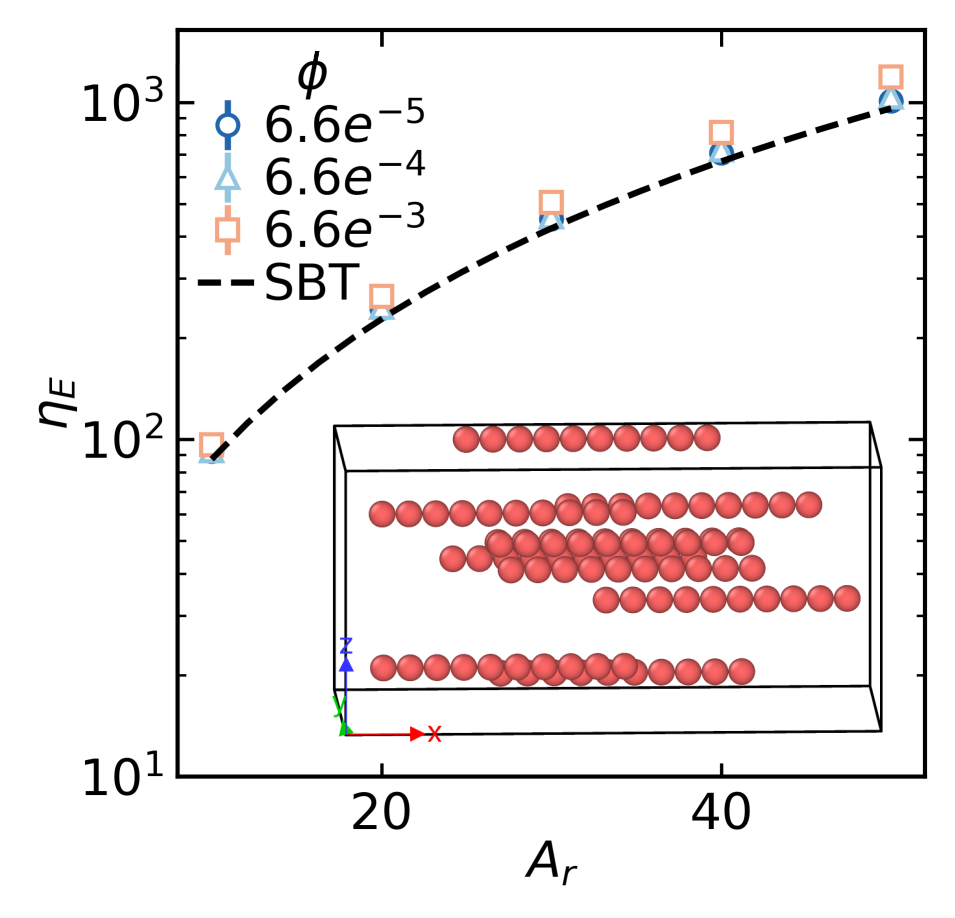}
\caption{Extensional viscosity of an aligned, dilute random fiber suspension as a function of fiber aspect ratio under uniaxial extensional flow. Results are compared with predictions from the modified slender-body theory~\cite{mackaplow1996numerical}.}
\label{fig:fiber_stress}
\end{figure}

We next compute the extensional viscosity of a dilute suspension of randomly distributed fibers aligned with the principal axis of uniaxial extensional flow. The viscosity is obtained by ensemble averaging over 10 independent configurations. Figure~\ref{fig:fiber_stress} shows the extensional viscosity as a function of aspect ratio $A_r$ for three volume fractions in the range $\phi = 6.6\times10^{-5}$ to $6.6\times10^{-3}$. The viscosity increases monotonically with increasing $A_r$, consistent with theoretical expectations. At the lowest volume fraction, $\phi = 6.6\times10^{-5}$, the results are in excellent agreement with the modified slender-body theory~\cite{mackaplow1996numerical}. At higher volume fractions, small deviations from theory emerge, which can be attributed to enhanced hydrodynamic interactions between fibers that are not captured in the dilute-limit theory.

\subsection{Method performance}
We benchmark the scaling performance of the FSD algorithm on single GPU \verb|NVIDIA A40| with $48GB$ RAM. The reported metric is average wall-clock runtime per timestep per bead, $t_s$, averaged over $100$ production timesteps after discarding the first 10 steps for equilibration. The simulation setup consists of $N$ Brownian rigid rods confined in a periodic cubic box, with each rod discretized into $n_b$ beads, giving a total of $N_{\rm tot}=N n_b$ beads. Rod–rod interactions are modeled via normal contact forces to prevent overlap. 

Because the computational cost depends on Ewald splitting parameter $\xi$, we consider two scenarios: (i) a fixed value of $\xi=0.5$, and (ii) a system-size–dependent value of $\xi=25/L$, to ensure consistent statistics across different box sizes. Scaling is analyzed as a function of $N_{\rm tot}$ at three volume fractions, $\phi \in \{0.01,0.05,0.10\}$ for two rod lengths, $n_b \in \{8,27\}$. 

\begin{figure}
\centering
\includegraphics[scale=1]{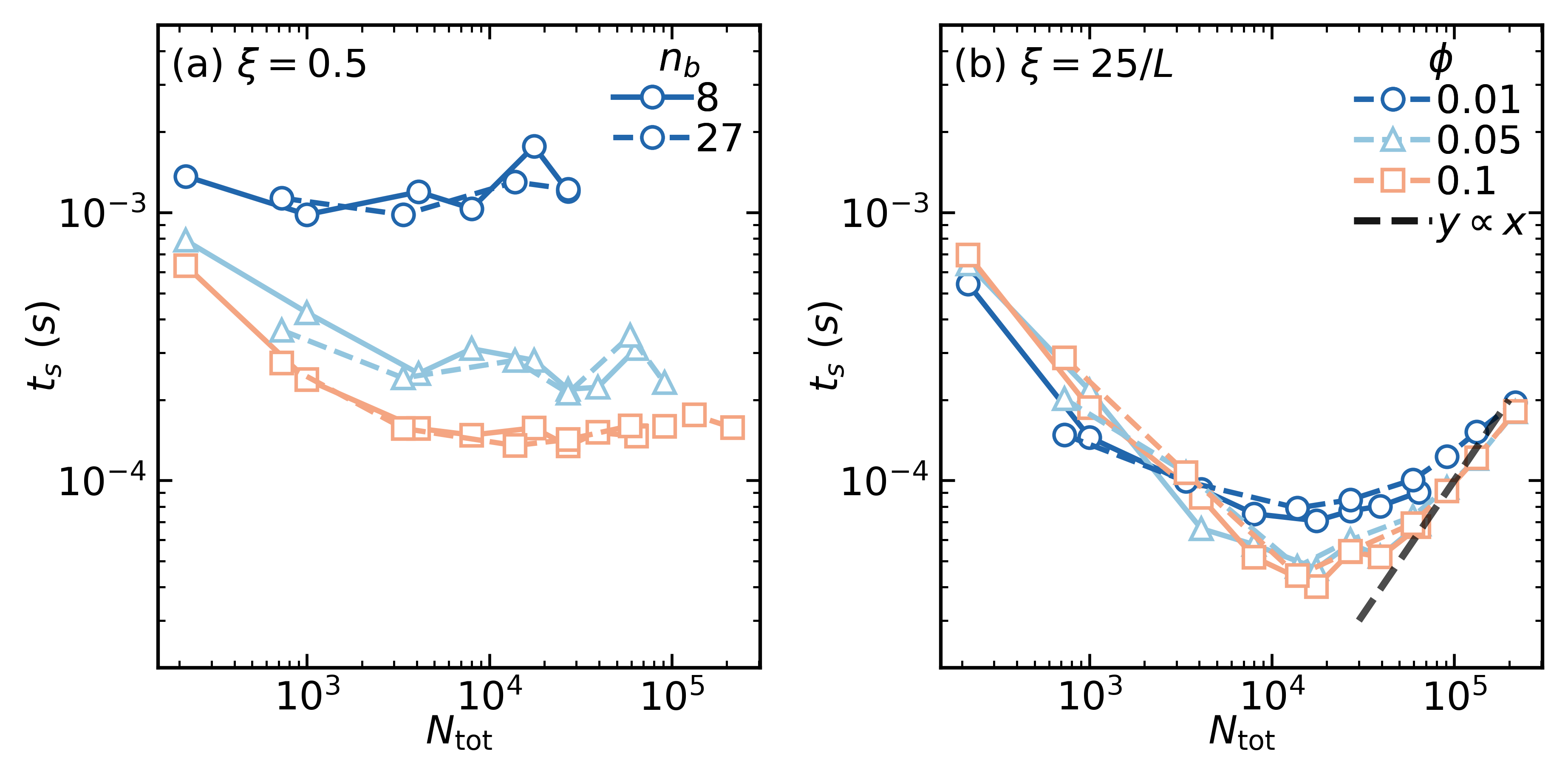}
\caption{Wall-clock runtime of the FSD algorithm as a function of total number of beads $N_{\rm tot}=Nn_b$ at different volume fractions $\phi$, where $N$ is the number of rigid rods and $n_b$ is the number of beads in a rod. Two cases are shown: (a) fixed Ewald splitting parameter $\xi=0.5$ (b) box-size dependent $\xi=25/L$. For comparison, scaling $t\propto x$ is also shown in (b).}
\label{fig:scaling}
\end{figure}

Figure~\ref{fig:scaling} reports the normalized run-time per bead, $t_s$, as a function of total number of beads $N_{\rm tot}$. For the fixed $\xi=0.5$, $t_s$ decreases with increasing $N_{\rm tot}$ before approaching a plateau at large $N_{\rm tot}$. The runtime also decreases with increasing $\phi$, and the data for both rod lengths collapse when plotted against $N_{\rm tot}$. In contrast, when using size-dependent $\xi=25/L$, $t_s$ initially decreases but exhibits an approximately linear increase with $N_{\rm tot}$ at large system sizes, with minimal dependence on $\phi$. Notably, at low $\phi$, simulations with size-dependent $\xi$ are almost an order of magnitude faster than those with fixed $\xi$. 

The initial decrease in $t_s$ at small $N_{\rm tot}$ arises from partial GPU memory usage. At large $N_{\rm tot}$, the GPU is fully utilized, and the asymptotic scaling reflects the algorithmic cost. The interplay between system size and $\xi$ can be understood through real- and reciprocal-space cutoffs, $R_c=\sqrt{-\log{x_{err}}}/\xi$ and $k_c=2\sqrt{-\log{x_{err}}}\xi+1$, respectively, where $x_{err}$ is the solver tolerance. 

For fixed-$\xi$ case ($\xi = 0.5$), both cutoffs remain constant, while the number of reciprocal vectors, $k = k_c L / \pi$, increases linearly with system size $L$. Consequently, for a fixed $N_{\rm tot}$, the total number of $k-$vectors scales as $k_{total}\propto L^3 \propto 1/\phi$, leading to higher computational cost at lower volume fractions. This scaling explains the observed order-of-magnitude difference in runtime between $\phi=0.01$ and $\phi=0.1$ at large system sizes. In contrast, for the size-dependent $\xi$ case ($\xi=25/L$), both $R_c$ and $k-{\rm vectors}$ increase linearly with system size $L$. This results in a growing computational burden with system size, consistent with the observed linear increase of $t_s$ at large $N_{\rm tot}$.

\section{Carbon black slurry rheology}\label{sec:CB}
The rheological behavior of carbon black (CB) slurries plays a critical role in the scalable fabrication of high-performance battery electrodes, since it directly influences how the conductive network forms during processing~\cite{shariq2024modelling}. Because CB particles tend to assemble into complex hierarchical structures driven by strong van der Waals attractions, the resulting slurries exhibit pronounced non-Newtonian behavior, including shear thinning and thixotropy~\cite{hipp2021direct}. These features make it challenging to balance processability during coating with the need to preserve a well-connected conductive pathway in the final electrode. As a result, accurate prediction and control of slurry viscosity under different shear conditions are essential to avoid defects such as particle sedimentation, non-uniform coating thickness, and drying-induced cracking, while still maintaining electrical connectivity. 

The microstructure of carbon black is inherently hierarchical spanning multiple length scales~\cite{hipp2021direct}. At the smallest scale, it consists of polydisperse, irregular primary particles of approximately $10-50$ nm, which may contain internal porosity depending on the CB type. These particles are irreversibly fused into primary aggregates, forming fractal structures of roughly $100-500$ nm that act as the fundamental building blocks of the suspension. At larger scales, these aggregates further assemble into $\upmu$m-sized agglomerates that exhibit shear-dependent restructuring under flow, strongly influencing the macroscopic rheology of the slurry.

\subsection{Generation of carbon black aggregates}\label{subsec:agg_modeling}
Inoue et al. \cite{inoue2016effect,inoue2019simulation} developed a numerical model to reconstruct realistic CB aggregate structures. This method is based on a probability density function,  
\begin{equation}
    \mathcal{P}(d_i) = 1 - (1-d_i)k_p,
    \label{eq:prob}
\end{equation}
where $\mathcal{P}(d_i)$ represents the probability of selecting particle $i$ as the connection site for next particle. The variable $d_i= L_i/\max(L_i)$ is the normalized distance sum, with $L_i$ being the total distance from particle $i$ to all other particles in the aggregate. The morphology parameter $k_p$ characterizes the strength of repulsive interactions; larger $k_p$ favors attachment to more distant particles, resulting in more elongated aggregate structures. The primary particle diameter and the number of particles per aggregate are specified as input parameters.

The aggregation process begins with a seed particle, to which a second particle is randomly attached. For third and subsequent particles, the distance sum for each existing particle is computed as $L_i=\sum_{j=1}^Nl_{ij}$, where $l_{ij}$ is the distance between particles $i$ and $j$. The probability function $\mathcal{P}(d)$ is then used to select candidate particles for attachment. Around the selected particles, multiple candidate positions for the new particle center are generated, allowing for slight overlap.  For each candidate position, the total distance to all previously placed particles is evaluated, and the corresponding probability is calculated using Eq.~\eqref{eq:prob}. The final placement is determined based on this probability distribution. This procedure is repeated until the desired number of particles are placed.

\begin{figure}
\centering
\includegraphics[scale=1.3]{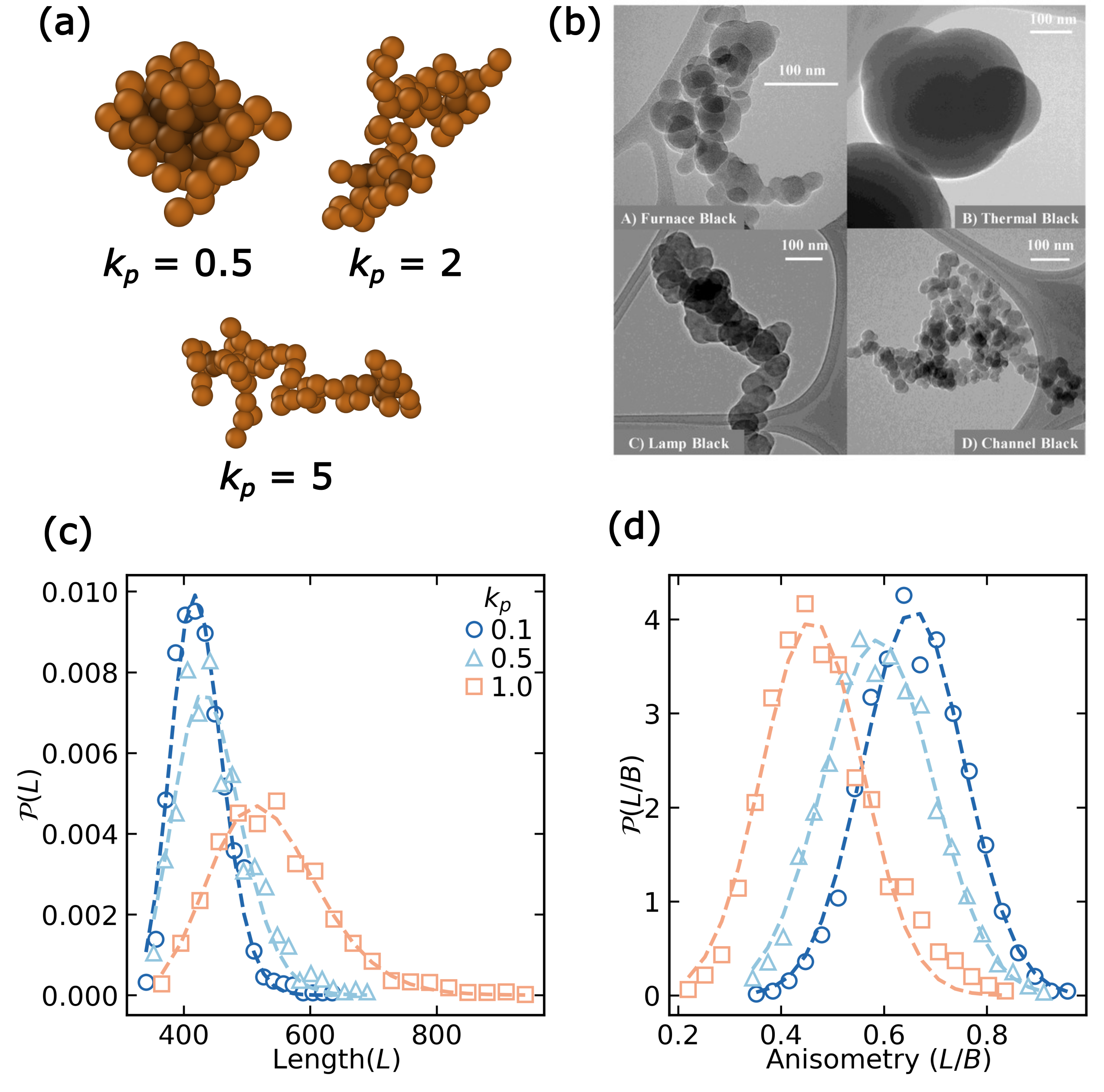}
\caption{Model structures of carbon black aggregates for (a) $k_p = 0.5$, 2, and 5, each consisting of $n_b = 64$ beads. (b) Representative TEM images of carbon black aggregates obtained from different sources \cite{singh2018nanostructure}. Statistical distributions of (c) aggregate length ($L$) and (d) anisometry ($L/B$) for varying morphology parameters. Dashed curves indicate lognormal (for $L$) and Gaussian (for $L/B$) fits, showing good agreement with experimental measurements \cite{grulke2018size}.}
\label{fig:aggregate}
\end{figure}

The aggregate structures shown in Fig.~\ref{fig:aggregate}(a) highlight the strong influence of the morphology parameter $k_p$ on the geometry of carbon black clusters. As $k_p$ increases from 0.5 to 5, the aggregates transition from relatively compact configurations to more elongated and anisotropic structures, for a fixed number of primary particles ($n_b = 64$). This behavior is consistent with the qualitative trends observed in the Transmission Electron Microscopy (TEM) images (Fig.~\ref{fig:aggregate}(b))~\cite{singh2018nanostructure}, which show significant variability in aggregate shape from different carbon black sources.

These morphological changes are further quantified through statistical analysis. The distribution of aggregate length ($L$) broadens and shifts toward larger values with increasing $k_p$, while the anisometry ($L/B$) exhibits a systematic increase, indicating enhanced shape anisotropy, where $B$ denotes the aggregate breadth (Fig.~\ref{fig:aggregate}(c,d)). The length distributions are well described by lognormal functions, reflecting the multiplicative nature of aggregation processes, whereas the anisometry follows an approximately Gaussian distribution. The good agreement between these fitted distributions and experimental measurements~\cite{grulke2018size} suggests that the model captures the essential geometric characteristics of real carbon black aggregates. These morphology-dependent variations are expected to play a critical role in determining the rheological behavior and network formation in suspensions. In particular, more elongated and anisotropic aggregates are likely to promote stronger mechanical interactions and facilitate the formation of percolated structures under flow.

\subsection{Interparticle interactions}\label{sec:interact}
The interactions between beads belonging to different carbon black aggregates are modeled by combining DLVO-type colloidal forces with a short-range mechanical contact description. The DLVO framework captures the interplay between van der Waals attraction and electrostatic double-layer repulsion, which together determine the stability and reversible aggregation behavior of particles in suspension. The DLVO potential between two particles of radii $a_1$ and $a_2$, separated by center-to-center distance $r$, is given by
\begin{equation}
    V_{\text{DLVO}}(r) = -\frac{A_H}{6}\left[\frac{2 a_1 a_2^2}{r^2-(a_1+a_2)^2}+\frac{2a_1 a_2}{r^2-(a_1-a_2)^2}+\ln\left(\frac{r^2-(a_1+a_2)^2}{r^2-(a_1-a_2)^2}\right)\right] + \frac{a_1 a_2}{r} Z e^{-\kappa(r-(a_1+a_2))},
    \label{eq:dlvo_potential}
\end{equation}
where $A_H$ is the Hamaker constant, $Z$ characterizes the surface electric potential, and $\kappa^{-1}$ is the Debye screening length. The corresponding DLVO force, $\mathbf{F}_{\text{DLVO}}$, is the spatial derivative of the interaction potential, given by
\begin{equation}
    \mathbf{F}_{\text{DLVO}}(r) = -\nabla V_{\text{DLVO}}(r).
\end{equation}

The van der Waals contribution diverges unphysically $\mathcal{O}(1/h^2)$ when surface-to-surface separation $h=r-(a_1+a_2)$ approaches zero. To regularize this behavior, we introduce a small cutoff length such that $h \rightarrow h+h_c$. For overlapping configuration ($h<0$), the attractive force is capped at its value at contact to avoid singular behavior.

To prevent the overlap at small separation, a normal contact force based on Hertz model, $\mathbf{F}_N=k_n\delta^{1.5}\mathbf{n}$ is introduced, where $k_n$ is the contact stiffness, $\delta=a_1+a_2-r$ is the overlap distance, $\mathbf{n}$ is the unit vector along the line of centers. 

Carbon black particles possess highly irregular and rough surfaces that can hinder relative sliding and rolling under shear, thereby strongly influencing suspension rheology~\cite{mari2014shear,wang2020hydrodynamic}. Two main approaches have been proposed to model this effect: (1) a stick-slip formulation based on Coulomb's friction law~\cite{mari2014shear}, and (2) a hydrodynamic friction framework via enhanced lubrication resistance. In this work, we adopt the latter approach. 

Instead of relying on the weak logarithmic divergence of standard sliding and rolling resistances for smooth spheres, the corresponding components of the hydrodynamic resistance tensor are modified to diverge algebraically as $\mathcal{O}(1/h)$ during close approaches, where $h=r-(a_1+a_2)$ is the surface-to-surface distance. Following Wang et al.~\cite{wang2020hydrodynamic}, we add a correction $f_{ij}^{\rm AB}$ to the smooth-sphere scalar resistance functions $Y_{ij}^{\rm AB}$. For particles with characteristic asperity height $h_0$ and roughness strength $\alpha^{\rm AB}$, the correction applied when $h<h_0$ is

\begin{equation}
    f_{ij}^{\rm AB}(h) = \alpha^{\rm AB} \frac{a}{h_0}\left(\frac{h_0}{h}-3\frac{h}{h_0}+2\frac{h^2}{h_0^2}\right)\mathcal{H}(h_0-h),
    \label{eq:tan_fric}
\end{equation}
where $a$ is the particle radius and $\mathcal{H}(h_0-h)$ is the Heaviside function. While Wang et al. assumed identical roughness strengths $\alpha$ for the FU, FE, and SE modes in the eq.~\ref{eq:tan_fric}, Majji et al.~\cite{majji2022parameterization} derived physically consistent ratios by enforcing mechanical constraints. Consistent with their analysis, we use $\alpha^{\rm FE}=\alpha^{\rm FU}/2$ and $\alpha^{\rm SE}=9\alpha^{\rm FU}/20$.

\subsection{Simulation Setup and shear protocol}
To assess the predictive capability of our framework against experimentally measured flow curves of carbon black slurries~\cite{bauland2024attractive}, we simulate dilute suspensions of the reconstructed fractal aggregates at volume fractions $\phi=0.024$ and $0.041$, under a transient ramp-down shear-rate protocol. The simulations are performed in a periodic cubic domain containing $N=125$ aggregates, each composed of $n_b=36$ primary beads of radius $a=20$ nm. We use morphological parameter $k_p=0.5$, yielding aggregates with radius of gyration $R_g=80$ nm and fractal dimension $d_f=2.35$, consistent with experimental observations. 

Because interparticle interactions are not directly characterized in the experiments, we adopt parameters from related literature~\cite{bauland2025antithixotropic}. Specifically, we use Hamaker constant $A_H=15k_B{T}$, zeta potential $Z=4k_B{T}$, and Debye screening length $\kappa^{-1}=20$ nm. The van der Waals interaction is regularized with a cutoff $h_c=0.6$ nm. The normal contact Hertz force is modeled using spring stiffness $k_n=5\times 10^5 k_B{T}/a^{2.5}$, ensuring that the maximum overlap remains below $5$\% of the bead radius throughout the simulation. Finally, surface roughness effects are incorporated via a tangential hydrodynamic friction model, with parameters $\alpha=0.01$ and $h_0/a=0.01$, corresponding to sub-nanometer roughness typical of graphitic layers. At the dilute volume fractions considered here, these parameters primarily influence short-range dissipation during near-contact interactions, while their impact on the structural evolution and overall stress is secondary compared to the contribution arising from constraint forces and DLVO interactions.

In the experiment, the samples were presheared at high shear rate, $\dot{\gamma}=1000s^{-1}$, to erase any prior shear history and ensure a rejuvenated state. Flow curves were then obtained by progressively decreasing the shear rate from $\dot{\gamma}=1000s^{-1}$ to $0.01s^{-1}$, using 10 points per decade, with durations  $\Delta{t}=1$, 50 and 100$s$ at each shear rate. The measured stress response was found independent of $\Delta{t}$ at high shear rates ($\dot{\gamma}>1s^{-1}$), while at low shear shear rates and high particle volume fractions, a pronounced time dependence was observed, attributed to anti-thixotropic dynamics~\cite{bauland2025antithixotropic}. 

To mimic these conditions, simulations are initialized from a well-dispersed configuration corresponding to the experimentally rejuvenated state. Directly matching experimental timescales is computationally prohibitive, particularly at high shear rates. However, given that the stress response is independent of duration for shear rates $\dot{\gamma}>1s^{-1}$, we adopt a reduced protocol in which the shear rate is ramped from $\dot{\gamma}=10^5s^{-1}$ to $10s^{-1}$, using 5 points per decade. At each shear rate, the system is evolved for a fixed accumulated strain of 4.

\subsection{Structure evolution and rheology under steady shear}
\begin{figure}
\centering
\includegraphics[scale=0.25]{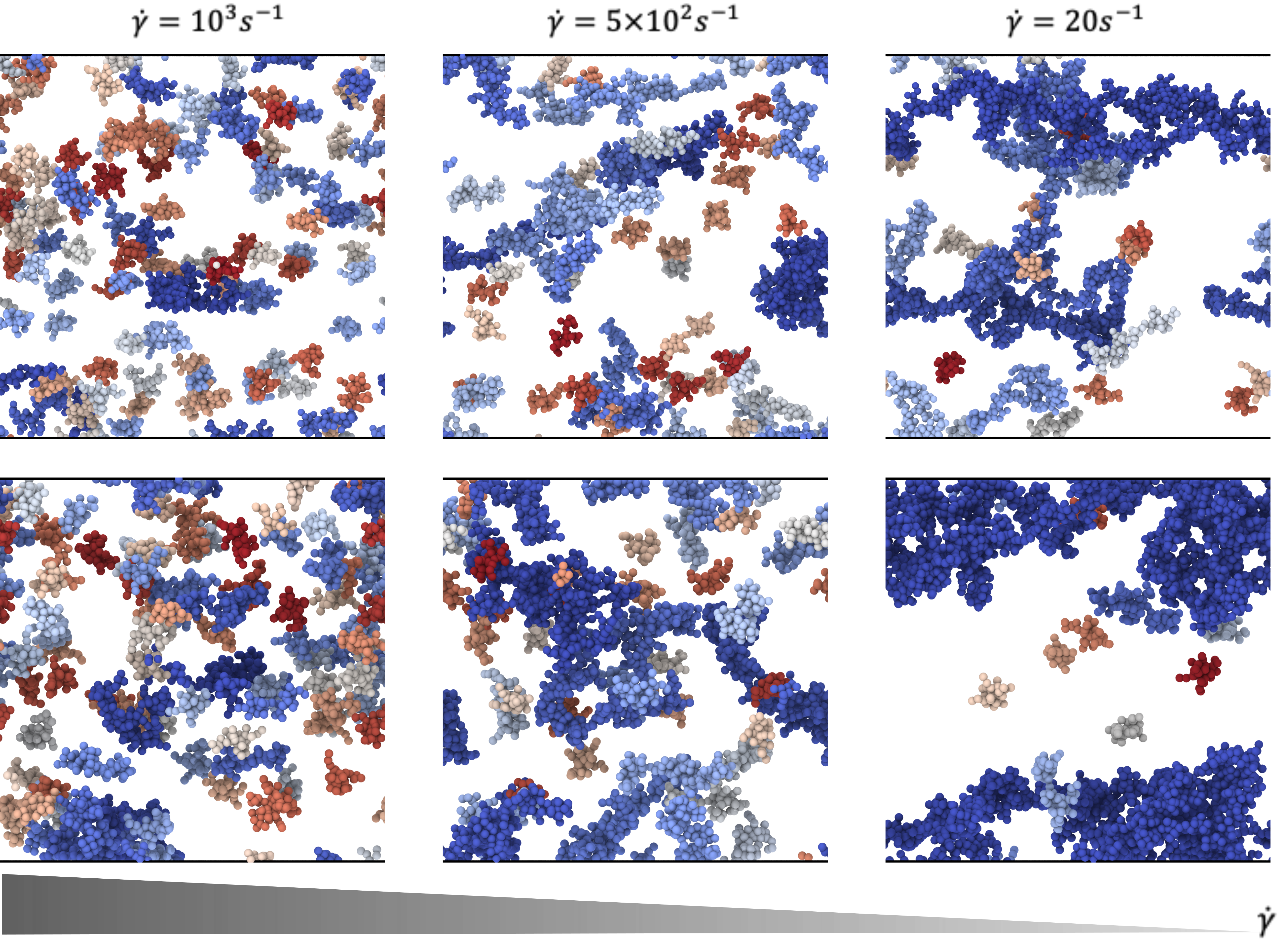}
\caption{Evolution of carbon black microstructure during ramp-down shear-rate flow for particle volume fractions $\phi = 0.024$ (top row) and $\phi = 0.041$ (bottom row). Particle beads are color-coded by their respective clusters.}
\label{fig:CBStructure}
\end{figure}

Figure~\ref{fig:CBStructure} shows representative microstructural snapshots at three shear rates: high $10^4s^{-1}$; intermediate $500s^{-1}$; and low $20s^{-1}$, for two volume fractions, $\phi=0.024$ (top row) and 0.041 (bottom row). The observed morphologies reflect a nonlinear competition between hydrodynamic stresses and attractive DLVO interactions.

At high shear rates, hydrodynamic forces dominate, maintaining a well-dispersed state for both volume fractions. The suspension is characterized by small, transient clusters. As the shear rate decreases to an intermediate value, attractive interactions become increasingly significant. The flow is no longer strong enough to disrupt particle contacts; instead, it promotes orthokinetic collisions between neighboring clusters. This leads to microstructural coarsening, where fractal aggregates merge into larger, anisotropic structures. At the lowest shear rate, the system undergoes a pronounced structural transition. Instead of discrete clusters, the suspension exhibits shear-induced phase separation. This effect is especially prominent at $\phi=0.041$, where the system separates into a dense, percolated particle network coexisting with a particle-depleted solvent region.

\begin{figure}
\centering
\includegraphics[scale=1.0]{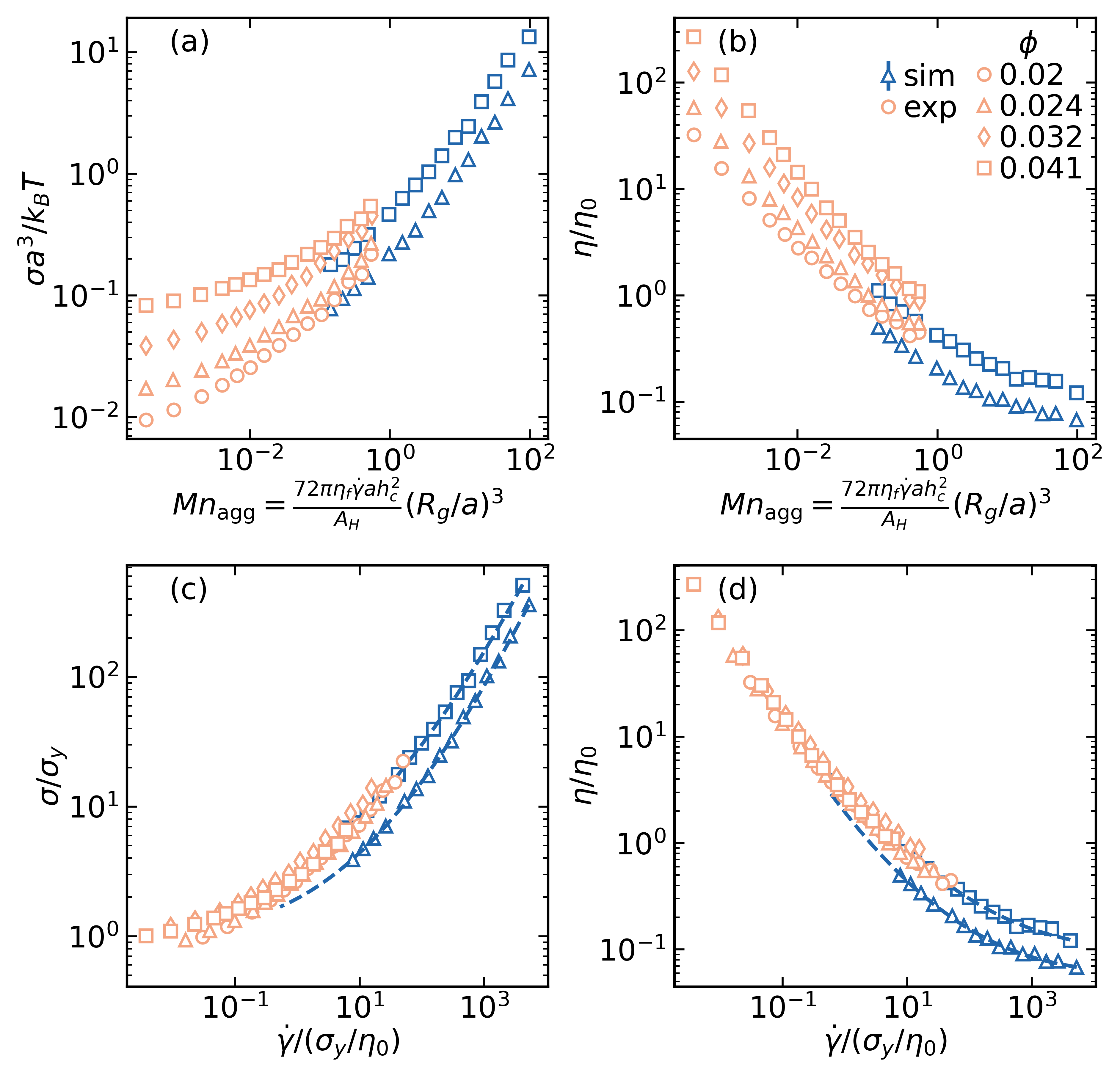}
\caption{Normalized flow curves of carbon black (CB) dispersions. Orange symbols show experimental data at volume fractions $\phi = 0.020$, 0.024, 0.032, and 0.041, while blue symbols show simulation results for $\phi = 0.024$ and 0.041. Top row (a,b): shear stress ($\sigma a^3/k_BT$) and relative viscosity ($\eta/\eta_0$) as functions of the Mason number, $Mn_{\rm agg}=72\pi\eta_f ah_c^2\dot{\gamma}(R_g/a)^3/A_H$. Bottom row (c,d): the same data rescaled by dynamic yield stress $\sigma_y$ and characteristic shear rate $\sigma_y/\eta_0$, leading to a collapse across all volume fractions for experiment data. Both quantities are obtained by fitting each flow curve to the Caggioni–Trappe–Spicer (CTS) model, shown by the dashed lines.}
\label{fig:CBflowcurves}
\end{figure}

We next evaluate the macroscopic steady-state flow curves and compare them with experiment. In Fig.~\ref{fig:CBflowcurves}(a,b), the data are normalized using the aggregate Mason number ($Mn_{\rm agg}=72\pi\eta_f ah_c^2\dot{\gamma}(R_g/a)^3/{A_H}$), which quantifies the ratio of hydrodynamic shear forces to interparticle attractive interactions at the aggregate scale. The experimental data (orange symbols) exhibit a characteristic yield-stress fluid response. At low Mason number ($<10^{-2}$), the stress approaches a plateau corresponding to the dynamic yield stress $\sigma_y$, and at higher shear rates the system displays pronounced shear-thinning behavior. The yield stress also increases systematically with particle volume fraction $\phi$, consistent with stronger network formation at higher loading. 

The simulation data (blue symbols) reproduce the shear-thinning regime at intermediate Mason numbers and the crossover toward a Newtonian-like response at large $Mn_{\rm agg}$, but they do not show the low-shear stress plateau seen experimentally. This is because simulations were restricted to $Mn_{\rm agg}>0.1$: at low shear rates the suspension undergoes shear-induced phase separation, preventing access to the homogeneous yielding regime. Consequently, within the accessible window, the simulated stresses continue to decrease via power-law scaling rather than leveling off. In addition, the simulation curves lie systematically below the experimental measurements. This offset likely reflects uncertainty in the model input parameters--particularly the interparticle attraction strength $A_H$, which was not directly measured, and ambiguity in the experimental estimate of $\phi$-- rather than a failure of the underlying physics.

To factor out these parameter uncertainties, Fig.~\ref{fig:CBflowcurves}(c,d) follows a self-consistent normalization used in experimental studies. Here, the data are non-dimensionlized using material-specific parameters extracted from the Caggioni–Trappe–Spicer (CTS) fit, namely the dynamic yield stress $\sigma_y$ and critical shear rate $\sigma_y/\eta_0$, where $\eta_0$ is the high-shear viscosity). The CTS model~\cite{bauland2024attractive}
\begin{equation}
    \sigma = \sigma_y + \sigma_y \left(\frac{\dot{\gamma}}{\dot{\gamma}^p}\right)^{1/2} + \eta_0 \dot{\gamma}
\end{equation}
captures the crossover from yielding to shear thinning and eventually to the high-shear viscous Newtonian regime. With this normalization, the experimental data across all volume fractions collapse onto a common master curve, indicating that the main rheological differences can be absorbed into $\sigma_y$ and $\eta_0$. 

The simulation data, however, remain systematically offset even after this rescaling. We attribute this discrepancy primarily to the absence of the low-shear yielding regime in the simulations rather than to the normalization procedure itself. Because the simulated flow curves contain only the shear-thinning and high-shear regimes, the CTS model must extrapolate to estimate the yield stress, resulting in considerable uncertainty in the fitted parameters. Consequently, the simulations do not sample the homogeneous yielding regime required to determine $\sigma_y$ reliably. Despite this limitation, the framework accurately captures the post-yield rheology across the shear-thinning and high-shear viscous regimes. Extending the method to access the homogeneous yielding regime remains an important direction for future work.

\section{Conclusions}
In this work, we implemented a fast Stokesian dynamics (FSD) framework for suspensions of rigid aggregates by extending the original sphere-level formulation of Fiore and Swan~\cite{fiore2019fast} to handle multi-bead rigid bodies. Rigidity is enforced through an implicit formulation based on geometric constraints, enabling stable and efficient simulation of aggregate dynamics without explicitly resolving constraint forces at each time step. To accelerate the solution of the resulting saddle-point system, we constructed a block-triangular preconditioner that combines an approximate inverse of the far-field mobility with a block-diagonal approximation of the Schur complement. For the Schur complement, we retain only the diagonal blocks associated with individual rigid aggregates and invert them independently using LU decomposition, thereby avoiding the need to assemble and factorize the full coupled system.

The proposed framework was rigorously validated against a diverse set of deterministic and stochastic benchmarks. These tests included the motion of a doublet in shear flow, pairwise sedimentation, and the translational and rotational diffusivities of aggregate suspensions, alongside stress predictions for both ordered and dilute systems. Across all evaluations, the method demonstrated excellent agreement with established theoretical and numerical results, confirming its accuracy in capturing both complex hydrodynamic interactions and Brownian fluctuations.

We further tested its predictive capability against experimentally measured flow curves of carbon black slurries, accounting for cohesion due to van der Waals, normal contact using a Hertz model, and tangential (frictional) contact via modified lubrication force. The simulations reproduce the shear-thinning and high-shear viscous regimes seen experimentally but do not capture the low-shear yielding plateau, largely because phase separation prevents access to the homogeneous yielding regime.

We also assessed the computational performance of the implementation. The scaling analysis shows that  the normalized runtime per bead decreases with system size before plateauing as GPU utilization saturates. For a fixed Ewald splitting parameter, the runtime depends strongly on volume fraction due to the scaling of reciprocal-space computations, whereas a size-dependent choice significantly accelerates simulations at low volume fractions, yielding up to an order-of-magnitude speedup. At larger system sizes, however, the size-dependent approach exhibits linear growth in runtime, reflecting the increasing cost of both real- and reciprocal-space evaluations.

Overall, this work extends fast Stokesian dynamics to complex rigid aggregates while preserving its scalable matrix-free formulation, providing an efficient framework for simulating Brownian suspensions of arbitrarily shaped particles. Future work will focus on multi-GPU parallelization and improved preconditioners that more accurately capture inter-aggregate hydrodynamic coupling, enabling simulations of larger, more concentrated systems relevant to industrial and biological applications.

\clearpage

\appendix
% --- Redefine Figure Labeling ---
\renewcommand{\thefigure}{\thesection.\arabic{figure}}
\setcounter{figure}{0}

% --- Redefine Table Labeling ---
\renewcommand{\thetable}{\thesection.\arabic{table}}
\setcounter{table}{0}

% --- Redefine Algorithm Labeling ---
% Note: Use the counter name used by your specific package (usually 'algorithm')
\renewcommand{\thealgorithm}{\thesection.\arabic{algorithm}}
\setcounter{algorithm}{0}

\section{Brownian drift}\label{app_sec:browniandrift}
In the FSD framework, the Brownian drift term in Eq.~\eqref{eq:euler2} is computed separately and subsequently added to the particle velocities obtained from the saddle-point solve in Eq.~\eqref{eq:SP2}. In Stokesian dynamics, two common numerical approaches are used to approximate this drift. 

The first approach is the Random Finite Difference (RFD) method, which explicitly approximates the divergence of the mobility. This is done by applying small, random perturbations to the particle configuration and forming a centered finite-difference approximation,

\begin{equation}
    \nabla \cdot \mathbf{M} = \frac{(\mathbf{M}(\mathbf{x}+\frac{\epsilon}{2}\Delta{\mathbf{q}}) - \mathbf{M}(\mathbf{x}-\frac{\epsilon}{2}\Delta{\mathbf{q}}))\cdot \Delta{\mathbf{q}}}{\epsilon},
    \label{eq:rfd}
\end{equation}
where $\Delta{\mathbf{q}}$ is a Gaussian random vector and $\epsilon$ is a small parameter controlling the perturbation size. When selecting the value of $\epsilon$, we must balance the truncation error of the RFD approximation $\mathcal{O}(\epsilon^2)$ with the relative accuracy in computing the mobility-vector product $\mathbf{M}\cdot\Delta{\mathbf{q}}$. In the present FSD implementation, the product is computed using iterative solver with relative accuracy $\delta_{GMRES}$. Following Ref.~\cite{sprinkle2017large}, the optimal error balancing is achieved by selecting $\epsilon \sim \delta_{GMRES}^{1/3}$.
In this method, two saddle-point systems are solved at perturbed configurations $\mathbf{x}\pm\frac{\epsilon}{2}\Delta{\mathbf{q}}$, yielding velocities $\mathbf{U}_{\pm}$ and stresses $\mathbf{S}_{\pm}$:
\begin{equation}
    \begin{pmatrix} \mathbf{M}^{\infty}\left(\mathbf{x}\pm\frac{\epsilon}{2}\Delta{\mathbf{q}}\right) & \mathcal{B} \\  \mathcal{B}^T &  -\mathbf{R}_{\rm FU}^{\rm nf}\left(\mathbf{x}\pm\frac{\epsilon}{2}\Delta{\mathbf{q}}\right) \end{pmatrix} \cdot \begin{pmatrix} \begin{pmatrix}
        \mathbf{F}^{\rm ff} \\ \mathbf{S}^{\rm ff}_{\pm}
    \end{pmatrix} \\ \mathbf{U}_{\pm} \end{pmatrix} = \begin{pmatrix} \mathbf{0} \\ -\Delta{\mathbf{q}}  \end{pmatrix},
\end{equation}
\begin{equation}
    \mathbf{S}_{\pm} = \mathbf{S}_{\pm}^{\rm ff}-\mathbf{R}_{\rm SU}^{\rm nf}\left(\mathbf{x}+\frac{\epsilon}{2}\Delta{\mathbf{q}}\right) \cdot \Sigma^T \cdot \mathbf{U}_{\pm}.    
\end{equation}
The Brownian drift is then obtained from centered differences:
\begin{equation}
    \mathbf{U}_{\rm drift} = k_B T \nabla \cdot \mathbf{R}^{-1}_{\rm FU} \approx k_B T \frac{\mathbf{U}_+ - \mathbf{U}_-}{\epsilon},    
\end{equation}

\begin{equation}
    \mathbf{S}^{\rm B} = k_B T \nabla \mathbf{R}_{\rm SU} \cdot \mathbf{R}^{-1}_{\rm FU} \approx k_B T \frac{\mathbf{S}_+ - \mathbf{S}_-}{\epsilon}    
\end{equation}

Another widely used to capture the drift term is modified mid-point method by Fixman\cite{fixman1978simulation}, which incorporates the Brownian drift implicitly through a weakly first-order accurate time discretization. In this method, Brownian increments are first generated using the mobility evaluated at the current configuration $\mathbf{q}$ through Lanczos-based method, 
\begin{equation}
    \begin{pmatrix}
    \mathbf{u}^{\rm B} \\ \mathbf{E}^{\rm B}
    \end{pmatrix}  = {\mathbf{M}^{\infty}(\mathbf{x})}^{1/2} \cdot \boldsymbol{\psi}^{\rm ff},
\end{equation}
\begin{equation}
    \mathbf{F}^{\rm B}_{\rm nf} = {\mathbf{R}_{\rm FU}^{\rm nf}(\mathbf{x})}^{1/2} \cdot \boldsymbol{\psi}^{\rm nf}.
\end{equation}
A saddle-point system is then solved to obtain provisional velocity $\mathbf{U}_0$ (and stress $\mathbf{S}_0$), 
\begin{equation}
    \begin{pmatrix} \mathbf{M}^{\infty}(\mathbf{x}) & \mathcal{B} \\  \mathcal{B}^T &  -\mathbf{R}_{\rm FU}^{\rm nf}(\mathbf{x}) \end{pmatrix} \cdot \begin{pmatrix} \begin{pmatrix}
        \mathbf{F}^{\rm ff} \\ \mathbf{S}^{\rm ff}
    \end{pmatrix} \\ \mathbf{U}_{0} \end{pmatrix} = \begin{pmatrix} \begin{pmatrix}
        \mathbf{u}^{\rm B} \\ \mathbf{E}^{\rm B}
    \end{pmatrix} \\ -\mathbf{F}^{\rm nf}_{\rm FU}  \end{pmatrix},    
\end{equation}

\begin{equation}
    \mathbf{S}_{0} = \mathbf{S}^{\rm ff}-\mathbf{R}_{\rm SU}^{\rm nf}(\mathbf{x})\cdot \Sigma^T \cdot \mathbf{U}_{0}.    
\end{equation}
The particle positions are advanced to an intermediate configuration $\mathbf{x}' = \mathbf{x} + \frac{\mathrm{d}{t}}{m}\mathbf{U}_0$, where $m$ is typically on the order of 100. Next, a second saddle-point system is solved to obtain updated velocity $\mathbf{U}'$ (and stress $\mathbf{S}'$), 
\begin{equation}
    \begin{pmatrix} \mathbf{M}^{\infty}(\mathbf{x'}) & \mathcal{B} \\  \mathcal{B}^T &  -\mathbf{R}_{\rm FU}^{\rm nf}(\mathbf{x'}) \end{pmatrix} \cdot \begin{pmatrix} \begin{pmatrix}
        \mathbf{F}^{\rm ff} \\ \mathbf{S}^{\rm ff}
    \end{pmatrix} \\ \mathbf{U}' \end{pmatrix} = \begin{pmatrix} \begin{pmatrix}
        \mathbf{u}^{\rm B} \\ \mathbf{E}^{\rm B}
    \end{pmatrix} \\ -\mathbf{F}^{\rm nf}_{\rm FU}  \end{pmatrix},    
\end{equation}

\begin{equation}
    \mathbf{S}' = \mathbf{S}^{\rm ff}-\mathbf{R}_{\rm SU}^{\rm nf}(\mathbf{x'}) \cdot \Sigma^T \cdot \mathbf{U}'.    
\end{equation}
The Brownian drift is then recovered from the difference between these two velocities (and stresses):
\begin{equation}
     \mathbf{U}_{\rm drift} = k_B T \nabla \cdot \mathbf{R}^{-1}_{\rm FU} \approx \frac{m}{2}(\mathbf{U}' - \mathbf{U}_0), 
\end{equation}

\begin{equation}
    \mathbf{S}^{\rm B} = k_B T \nabla \mathbf{R}_{\rm SU} \cdot \mathbf{R}^{-1}_{\rm FU} \approx -\frac{m}{2}(\mathbf{S}' - \mathbf{S}_0).
\end{equation}

Both RFD and Fixman schemes are formally consistent with It\^o stochastic calculus and recover the correct equilibrium statistics. However, they differ in several aspects including numerical cost, variance properties, and sensitivity to solver tolerances, particularly when the mobility operator is applied approximately using iterative or matrix-free methods.

\begin{figure}
\centering
\includegraphics[scale=1]{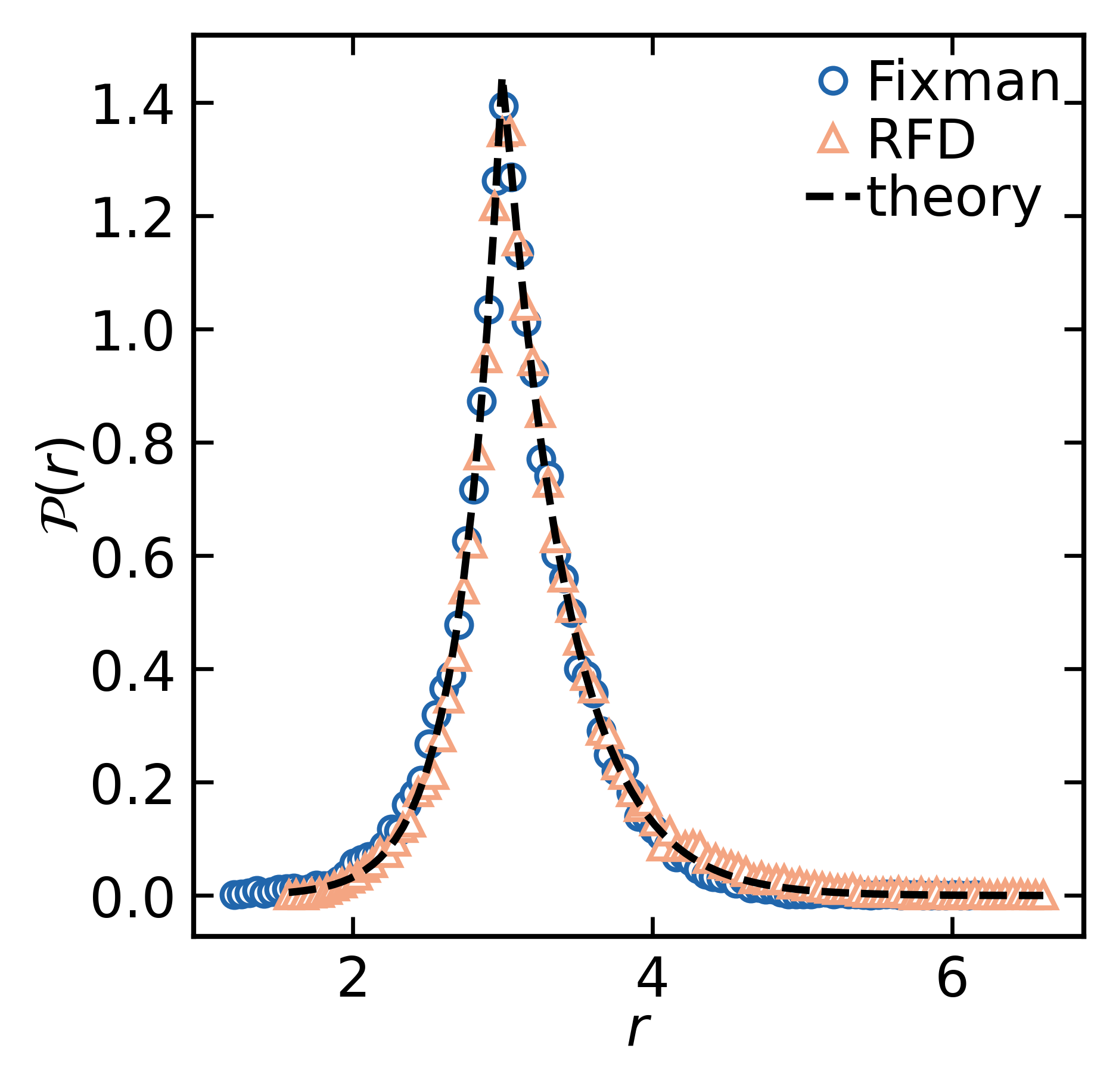}
\caption{Probability distribution $\mathcal{P}(r)$ for the center-to-center distance of two particles connected by a linear spring with spring constant $k_n=3k_B{T}$ and equilibrium separation $r_0=3$. Symbols denote simulation results obtained using the Fixman and RFD methods, while the dashed line represents the theoretical Boltzmann distribution.}
\label{fig:browniandrift}
\end{figure}

Both approaches were validated using a system of two particles interacting through a linear potential $V(r)=-k_n|r-r_0|$ with spring constant $k_n=3k_B{T}$ and equilibrium separation $r_0=3$. The resulting probability distribution $\mathcal{P}(r)$ for the center-to-center distance $r$ obtained from both the Fixman and RFD methods shows excellent agreement with the theoretical Boltzmann distribution (Fig.~\ref{fig:browniandrift}).

\clearpage

\bibliographystyle{elsarticle-num}
\bibliography{manuscript}
\end{document}